\documentclass[11pt,fleqn]{article}
\usepackage{epsfig}
\usepackage{subfigure}
\begin{document}

\def\lsim{\mathrel{\rlap{\lower4pt\hbox{\hskip1pt$\sim$}}
    \raise1pt\hbox{$<$}}}         
\def\gsim{\mathrel{\rlap{\lower4pt\hbox{\hskip1pt$\sim$}}
    \raise1pt\hbox{$>$}}}         
\def\esim{\mathrel{\rlap{\raise2pt\hbox{$\sim$}}
    \lower1pt\hbox{$-$}}}         

\hyphenation{neu-tral-ino neu-tral-inos higgs-ino higgs-inos gaug-ino
  gaug-inos}

\newcommand{\beq}{\begin{equation}}
\newcommand{\eeq}{\end{equation}}
\newcommand{\beqa}{\begin{eqnarray}}
\newcommand{\eeqa}{\end{eqnarray}}
\begin{titlepage}
\pagestyle{empty}
\title{Observability of $\gamma$ Rays from  Dark Matter \\
Neutralino
Annihilations in the Milky Way Halo}
\author{{\bf Lars Bergstr{\"o}m}\thanks{lbe@physto.se},  
 {\bf Piero Ullio}\thanks{piero@physto.se} \\
 Department of Physics, Stockholm University,\\
 Box 6730, SE-113~85~Stockholm, Sweden
\and
 {\bf James H. Buckley}\thanks{buckley@wuphys.wustl.edu}\\
Department of Physics\\
Washington University\\
St. Louis, MO 63130, USA}

\date{December 23, 1997}

\maketitle
\smallskip

\begin{abstract}
Recent advances in N-body simulations of cold dark matter halos point to
a substantial density enhancement near the center. This means that, e.g.,
the $\gamma$ ray signals from neutralino dark matter annihilations
would be significantly enhanced compared to old estimates based on
an isothermal sphere model with large core radius.

Another important development concerns new detectors, both space- and
ground-based, which will cover the window between 50 and 300 GeV 
where presently no cosmic $\gamma$-ray data are available.

Thirdly, new calculations of the $\gamma$-ray line signal (a sharp
spike of $10^{-3}$ relative width) from neutralino annihilations
have revealed a hitherto neglected contribution which, for heavy
higgsino-like neutralinos, gives an annihilation rate an order of
magnitude larger than previously predicted.

We make a detailed phenomenological study of the possible detection
rates given these three pieces of new information. We show that 
the proposed upgrade of the  Whipple telescope will make it sensitive to
a region of parameter space, with substantial improvements possible
with the planned new generation of Air Cherenkov Telescope Arrays.
We also comment on the potential of the GLAST satellite detector.

An evaluation of the continuum $\gamma$-rays produced in neutralino
annihilations into the main modes is also done. It is shown that a
combination of high-rate models and very peaked halo models are already
severely constrained by existing data.

\end{abstract}
\end{titlepage}

\section{Introduction}\label{sec:intro}

The dark matter problem is one of the most outstanding problems
confronting cosmology and astrophysics today. 
The question of the nature of the dark matter in the Universe is 
a cross-disciplinary one which may need elements of particle
physics for its solution. With new observations, the possible candidates 
get more and more constrained. 

Let us first recall that from the particle physics point of view, the
theoretically preferred (Einstein-De Sitter) 
Universe has the simple description 
\beq
\left\{
\begin{array}{ccc}
	\Omega_{tot} & = & 1  \\
	\Omega_\Lambda & = & 0\label{eq:eq1}
\end{array}\right.
\eeq
where as usual we have introduced the normalization of the energy density to the critical density 
\beq
\Omega\equiv {\rho\over \rho_{crit}}={\rho\over 1.9\cdot 10^{-29}h^2\ 
{\rm g}\,{\rm cm^{-3}}},
\eeq
and $\Omega_\Lambda$ is the contribution to the energy density from
a cosmological constant (or equivalently from vacuum energy).

The Einstein--De Sitter model (\ref{eq:eq1}) has the attractive features that 
it is simple, avoids finetuning, and may be explained by a period of  
inflation in the earliest Universe.
Since Big Bang nucleosynthesis (BBN) puts an upper limit to
 the  baryonic contribution  $\Omega_{b}$ 
of  \cite{olive}
\beq
\Omega_{b}h^2\leq 0.026\; ,\label{eq:bbn}
\eeq
with $h$ related to the Hubble constant $H_{0}$ by $h=H_{0}/(100\ {\rm 
km}\,{\rm s}^{-1}{\rm Mpc}^{-1})$ (observationally, $h$ lies between 
0.4 and 0.8),
non-baryonic dark matter dominates the energy density by a large 
factor in this type of model. Indeed, there is a number of different 
observations on scales from dwarf galaxies and upwards which point to
a larger value of $\Omega_{tot}$ than that allowed by nucleosynthesis
\cite{lbetaup}. It therefore seems that non-baryonic dark matter has
to be present in substantial amounts in the Universe today.

One of the prime 
candidates for the non-baryonic component is provided by the lightest 
supersymmetric particle, plausibly the lightest neutralino 
$\chi$~\cite{lbetaup}.
Supersymmetry seems to be a necessity in superstring theory (or 
M-theory) which unites all the 
fundamental forces of nature, including gravity. In most versions of 
the low-energy theory  there is a conserved multiplicative quantum 
number, R-parity, which makes the lightest supersymmetric particle 
stable. Thus, pair-produced neutralinos in the early Universe which 
left thermal equilibrium as the Universe expanded should have a non-zero 
relic abundance today. If the scale of supersymmetry breaking is related to 
that of electroweak breaking, $\Omega_{\chi}$ may come out in the right order 
of magnitude to explain the non-baryonic dark matter. 

 In addition, neutralinos are generically found to decouple at a temperature 
 that is roughly $M_{\chi}/20$, which means that they are 
 non-relativistic already at decoupling and certainly behave as 
Cold Dark Matter (CDM) 
 by the time of matter dominance and structure formation. Most analyses of
large scale structure formation in the Universe indeed find that CDM models
(perhaps with a small addition of, e.g., massive neutrinos) give the
best description of observational data.

Although accelerators like the Tevatron at Fermilab and LEP at CERN have
already started to probe regions of parameter space, so far without finding
any supersymmetric signals, the constraints imposed are not yet very 
restrictive. Adding the requirement that neutralinos make up most
of the dark matter, one finds that there are viable models in the mass
interval 30 GeV $\lsim M_\chi\lsim$ 10 TeV. In the low mass range, the
direct detection at accelerators or in terrestrial detectors sensitive
to the weak interactions generated by $\chi$ particles of the Milky Way
halo as they pass the Earth are probably the most promising methods of 
detection (see \cite{jkg} for an extensive review).

At the high mass end (or, depending on halo paramaters, maybe even 
over the full mass range) indirect detection methods are competitive. 
The best indirect signals are given by neutrinos from the Sun or the central
region of the Earth, and by almost monoenergetic $\gamma$ rays lines
originating from neutralino annihilations in
the Milky Way halo.

Neutrinos can escape from the centre  
of the Sun or Earth, where 
neutralinos may have been gravitationally trapped and therefore their density 
enhanced. Gamma rays may result from loop-induced 
annihilations 
\mbox{$\chi\chi\to\gamma\gamma$~\cite{lp}} or $\chi\chi\to Z\gamma$
\cite{lp2}.

The rates of these processes are difficult to estimate because of 
uncertainties in 
the supersymmetric parameters, cross sections and halo density profile. 
However, in contrast to other proposed detection methods they have 
the virtue of giving  very 
distinct, ``smoking gun'' signals: high-energy neutrinos from the 
centre of the Earth or 
Sun, or monoenergetic photons with $E_\gamma = M_\chi$ or $E_\gamma = M_\chi
(1-m_{Z}^2/4\,M_{\chi}^2)$ from the halo. The neutrino signal has been 
thoroughly discussed in the literature \cite{neutrinos}. In this article,
we concentrate on $\gamma$ ray lines, where three recent lines of development
have prompted us to take a new look at this process (for early studies,
see \cite{oldlines}). 

Firstly, a number of numerical studies using N-body
simulations \cite{carlberg,navarro,kravtsov,moore} 
have shown that in CDM cosmologies the hierarchical
way that halos form means that they should have a density profile which
is quite steep with  a singular behaviour $\sim r^{-\alpha}$ with
estimates of $\alpha$ ranging from 0.2 to 1.7. The physics underlying this
behaviour is simple: in CDM models with scale-invariant gaussian initial
fluctuations small clumps go nonlinear first, where the large average
background density at the early epochs means that the smallest clumps are
densest. Through the amplifying action of gravity, larger regions of 
overdensity form by the merging of smaller substructures. Through tidal 
interactions the overall halo distribution gets smeared out, but near the 
center some of the densest clumps leave a trace in form of their high 
average overdensity. Taking the interplay between the CDM and baryonic
parts of the halo into account, a satisfactory description of the rotation
curves of even dwarf spheroidal galaxies can be achieved \cite{bs}.

Since the probability for two CDM particles such as
neutralinos to meet and annihilate each other is proportional to the square
of the neutralino density, this enhancement near the galactic center could
be of utmost importance for the $\gamma$ ray line signal.

The second development concerns the calculation of the processes
\mbox{$\chi\chi\to\gamma\gamma$} and $\chi\chi\to Z\gamma$, which only recently
were computed fully to one loop in the MSSM, the 
minimal supersymmetric extension of the Standard Model of particle physics
\cite{lp,lp2}. In these calculations it was shown that previously neglected
contributions could increase the predicted rates by an order of magnitude,
especially at the high-mass end of the allowed $M_\chi$ range.

In the high mass range, $M_\chi\gsim$ 250 GeV, existing Air Cherenkov 
Telescopes (ACT) could already be sensitive to the $\gamma$ ray lines. However,
the third interesting development is the planning and construction of a 
new generation of ACTs which will have larger area, lower threshold and
better energy resolution. This will improve substantially the discovery
potential of these telescopes for supersymmetric dark matter.
Also, plans are being made for a satellite-borne detector (GLAST 
is one of the design concepts) which can compensate a factor
$\sim 10^4$ smaller area than ACTs
by a larger angular acceptance, better energy resolution, lower energy 
threshold and longer integration time.

The plan of this article is as follows. In Section~\ref{sec:mssm}, we
give an introduction to the MSSM and the predictions of the $\gamma$ line
flux. In Section~\ref{sec:halo} we discuss various halo models and
their implications for the $\gamma$ ray flux in ACTs from neutralino
annihilations. In Section~\ref{sec:planned} we give an overview of present and
planned detectors for cosmic $\gamma$ rays, and in Section~\ref{sec:actdet}
we discuss the detection potential of several existing or upcoming ACTs.
In Section~\ref{sec:glast} we discuss the discovery 
potential of satellite detectors, and in Section~\ref{sec:cont}
we comment also on the diffuse continuum $\gamma$ rays that would be
produced as secondary annihilation products if neutralinos make up 
a significant part of the dark
matter. Finally, in Section~\ref{sec:disc} we discuss our results and give
some concluding remarks.

\section{Overview of MSSM Results for $\gamma$ Lines}\label{sec:mssm}

We have performed a detailed phenomenological analysis of the 
annihilation rate of non relativistic neutralinos into two photons 
and into a photon and a $Z$ boson. The two processes, which have the 
experimentally significant feature 
 of giving nearly monochromatic photons in the final state, 
 with an energy $E_{\gamma} \simeq M_{\chi}$ and 
$E_{\gamma} \simeq M_{\chi} (1 - M_Z^2/4 M_{\chi}^2)$, respectively, have been 
studied in Refs.\,~\cite{lp} and ~\cite{lp2} where for the first time 
a full one loop calculation of the two cross sections was performed.
Considering a broad selection of MSSM models in which the neutralino 
is the lightest supersymmetric particle and a good dark matter 
candidate, the aim of this Section is to examine in which cases 
the two annihilation processes are important. We will compare the 
relative values of the cross sections in view of extracting 
information on the nature of the neutralino from a possible detection 
of one or both the gamma ray lines from neutralino annihilations in 
the galactic halo.

In the minimal $N=1$ supersymmetric extension of the Standard Model 
four neutral spin-1/2 Majorana particles are introduced, the partners
of the neutral 
gauge bosons $\tilde{B}$, $\tilde{W}^3$ and the neutral CP-even higgsinos  
$\tilde{H}^0_1$, $\tilde{H}^0_2$. Diagonalizing the corresponding 
mass matrix, four mass eigenstates are obtained:
\beq
  \tilde{\chi}^0_i = 
  N_{i1} \tilde{B} + N_{i2} \tilde{W}^3 + 
  N_{i3} \tilde{H}^0_1 + N_{i4} \tilde{H}^0_2 
\eeq
The lightest of these, $\tilde{\chi}^0_1$ or simply $\chi$, is 
commonly referred as the neutralino. It is useful to introduce the 
gaugino fraction  $Z_g$ defined as
\beq
  Z_g = |N_{11}|^2 + |N_{12}|^2
\eeq
and classify  the neutralino as higgsino-like when $Z_g<0.01$, mixed 
when $0.01 \leq Z_g \leq 0.99$ and gaugino like if  $Z_g>0.99$.

The general R-parity conserving version of the MSSM is defined by 
63 free parameters~\cite{jkg}. Some simplifying assumptions are 
necessary to reduce the number of parameters and get to a format 
which is  numerically tractable (for a complete discussion of this 
procedure see Ref.~\cite{bg,joakim,jkg}). In the scheme we use there 
are seven free parameters: the higgsino mass parameter $\mu$, the 
gaugino mass parameter\footnote{The usual Grand Unification Theory 
(GUT) relations for gaugino masses are assumed in all the results
 presented, their release is not expected to give major modifications}
 $M_2$, the ratio $\tan \beta$ of the vacuum expectation values of the
 two Higgs fields $H^0_2$ and $H^0_1$, the mass of the pseudoscalar
 Higgs  $m_A$, the sfermion mass parameter $m_0$ and other two soft
 supersymmetry breaking parameters $A_b$ and $A_t$. Our analysis 
is based on 16 scans over this parameter space. Some of them are 
broad samplings in the parameters, others are dedicated to more 
specific portions of the parameter space that are interesting 
for the present purpose. Table~\ref{tab:par} contains the maximal and minimal
 values for each parameter; in the given intervals, or in most cases
 in subintervals, $\mu$ and $M_2$ are scanned logarithmically, 
$\tan \beta$ is mostly scanned linearly and in a few scans 
logarithmically, while all the other parameters are varied along 
a linear scale. Rejecting those models that violate all known 
experimental bounds, we are left with a sample of about 62,500 
supersymmetric models which could describe Nature. For each of 
these, the neutralino relic density $\Omega_\chi h^2$ has been computed
 using the exact treatment of resonances~\cite{paolograc} and 
including coannihilations between neutralinos and 
charginos~\cite{joakimpaolo}. We restrict to those models which 
can give the neutralino as the main constituent of galactic halos,
 selecting the models which have a relic density that is in the 
interval $0.025<\Omega_\chi h^2<1$ (the upper limit is given by the 
condition of not overclosing the universe, or equivalently giving a 
too low age of the Universe). About 22,000 models 
in our sample fulfill this condition and this is the selected sample
 that is shown in the graphs in this paper.

\begin{table} [t]
\begin{center}
$1000\; \rm{GeV}\;<\;|\mu|\;<\;30000\; \rm{GeV}$ \\
$1000\; \rm{GeV}\;<\;|M_{2} |\;<\;50000\; \rm{GeV}$ \\
$1.2\;<\;\tan \beta\;<\;50$ \\
$0 \; \rm{GeV}\;<\;m_A\;<\; 10000\; \rm{GeV}$ \\
$100 \; \rm{GeV}\;<\;m_0\;<\; 30000 \; \rm{GeV}$ \\
$-3\,m_0\;<\;A_b\;<\;3\,m_0$ \\
$-3\,m_0\;<\;A_t\;<\;3\,m_0$ 
\end{center}
\caption{Considered range for the free parameters in our realization 
of the Minimal Supersymmetric Standard Model.} 
\label{tab:par}
\end{table}

\begin{figure}[htb]
 \epsfig{figure=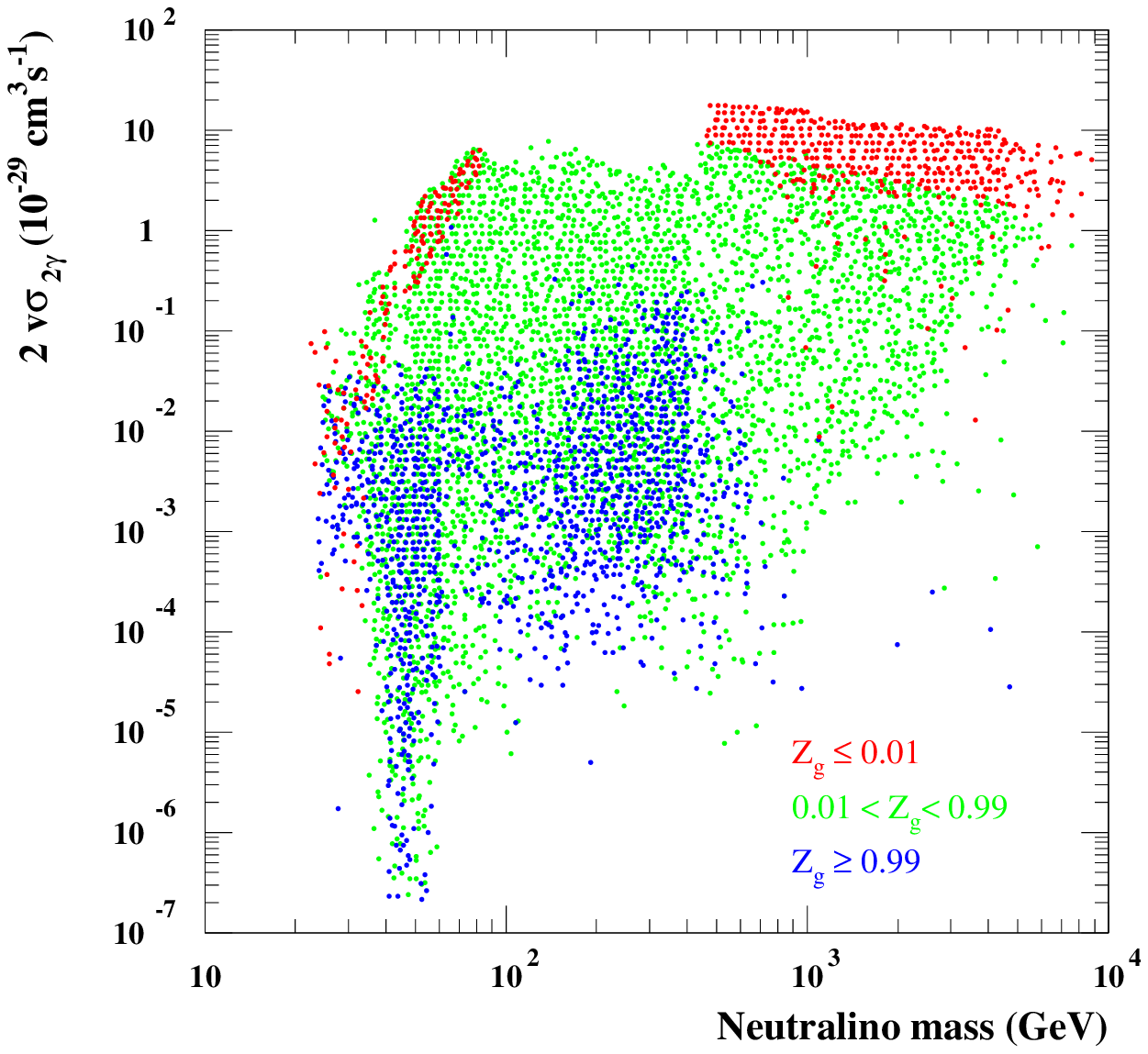,width=12cm}
\caption[]{Annihilation rate of neutralinos into the $2\gamma$ final 
state for the sample of supersymmetric models described 
in Section~\ref{sec:mssm}. A different marker color is used for the 
three classes of neutralinos we have defined: a red marker indicates 
a higgsino-like neutralino, a green marker a mixed neutralino, while 
blue markers are for gaugino-like neutralinos.}  
\label{fig:2gxs}
\end{figure}

\begin{figure}[htb]
 \epsfig{figure=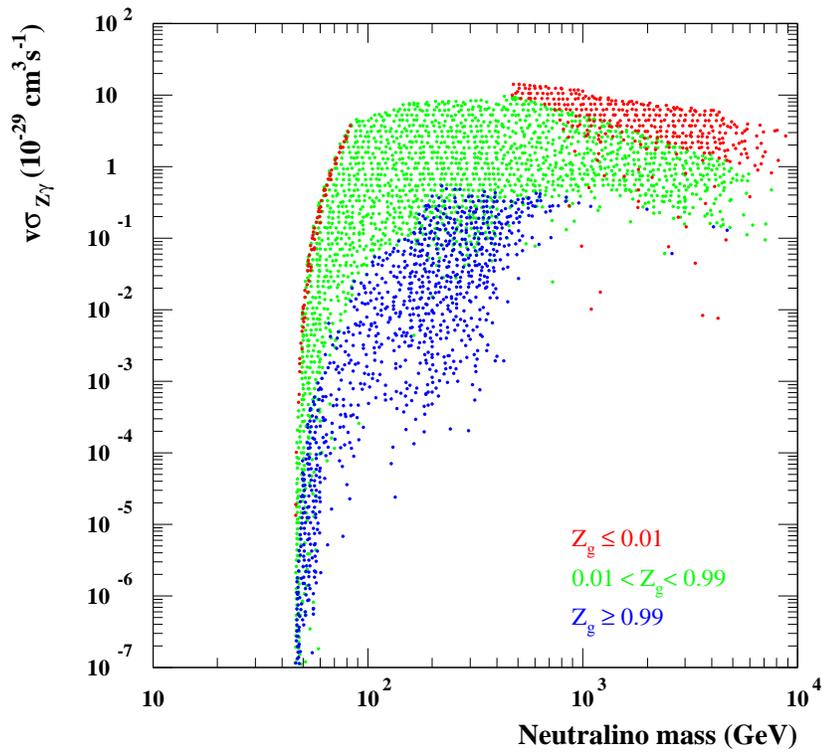,width=12cm}
\caption[]{Annihilation rate of neutralinos into the $Z\gamma$ final state
for the sample of supersymmetric models described in Section~\ref{sec:mssm}. 
The marker colors have the same meaning as in Fig.~\ref{fig:2gxs}.}
\label{fig:zgxs}
\end{figure}

\begin{figure}[htb]
 \epsfig{figure=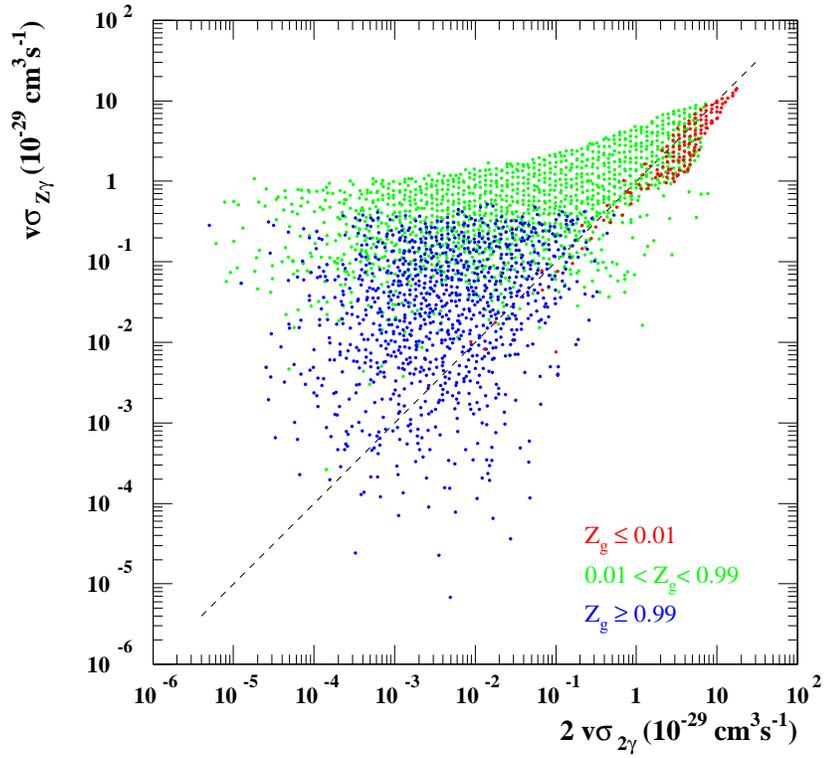,width=12cm}
\caption[]{$v \sigma_{Z\gamma}$ versus $2\,v \sigma_{2\gamma}$ for 
the sample of supersymmetric models described in Section~\ref{sec:mssm}. 
The marker color has the same meaning as in Fig.~\ref{fig:2gxs}.}
\label{fig:zgx2gx}
\end{figure}

In Fig.\,~\ref{fig:2gxs} and Fig.\,~\ref{fig:zgxs} we show 
the predictions for 
the annihilation rate of neutralinos into two photons and into a 
photon and a $Z$ boson as a function of the neutralino mass, while in 
Fig.\,~\ref{fig:zgx2gx} they are shown one versus the other. It is useful 
to compare $2\, v \sigma_{2\gamma}$ and $v \sigma_{Z\gamma}$ as the
 $\gamma$ line flux is proportional to these quantities (the 2 comes
 from the fact that there are two photons in that final state). 
In these figures as in all those that follow, each marker represents 
a model in our sample, we have however arbitrarily fixed a maximal 
density of points which are printed (it might be worth reminding 
that the point density in these kind of figures is meaningless and 
just depends on the way the parameter space is sampled). The visual
 effect is the same as drawing shaded zones but in this way we avoid
 the problem of deciding whether or not to include isolated points or 
 empty sub-regions.

A common feature of $2\, v \sigma_{2\gamma}$ and $v \sigma_{Z\gamma}$ is 
that the highest values are given by very pure higgsino-like 
neutralinos (we have included models with $Z_g$ as small as 
$10^{-6}$), with a mass around 500 GeV. As it was already shown 
in Refs.\,~\cite{lp} and ~\cite{lp2}, the cross sections tend to 
a constant value of about $10^{-28}\; \rm{cm}^{3}\,\rm{s}^{-1}$ 
in the limit $Z_g\to 0$ and $M_{\chi} \gg M_W$. The higgsino-like 
neutralino branch starts at very high masses, where the relic density 
is close to 1 (coannihilations of neutralinos with the lightest 
chargino are crucial to show that neutralinos with masses highier 
than 3 TeV can be cosmologically acceptable ~\cite{joakimpaolo}), 
and continues with increasing cross sections down to about 500 GeV 
where $\Omega_\chi h^2$ approaches 0.025. This is mainly due to the large
 cross section for the annihilation into a W boson pair. In fact, 
when this process becomes kinematically forbidden 
(for neutralino masses below the mass of the W boson) higgsino-like neutralinos 
become again cosmologically interesting candidates and, together 
with mixed neutralinos, again give the highest cross sections in 
that mass region. 

\begin{figure}
 \centering
 \mbox{{\epsfig{file=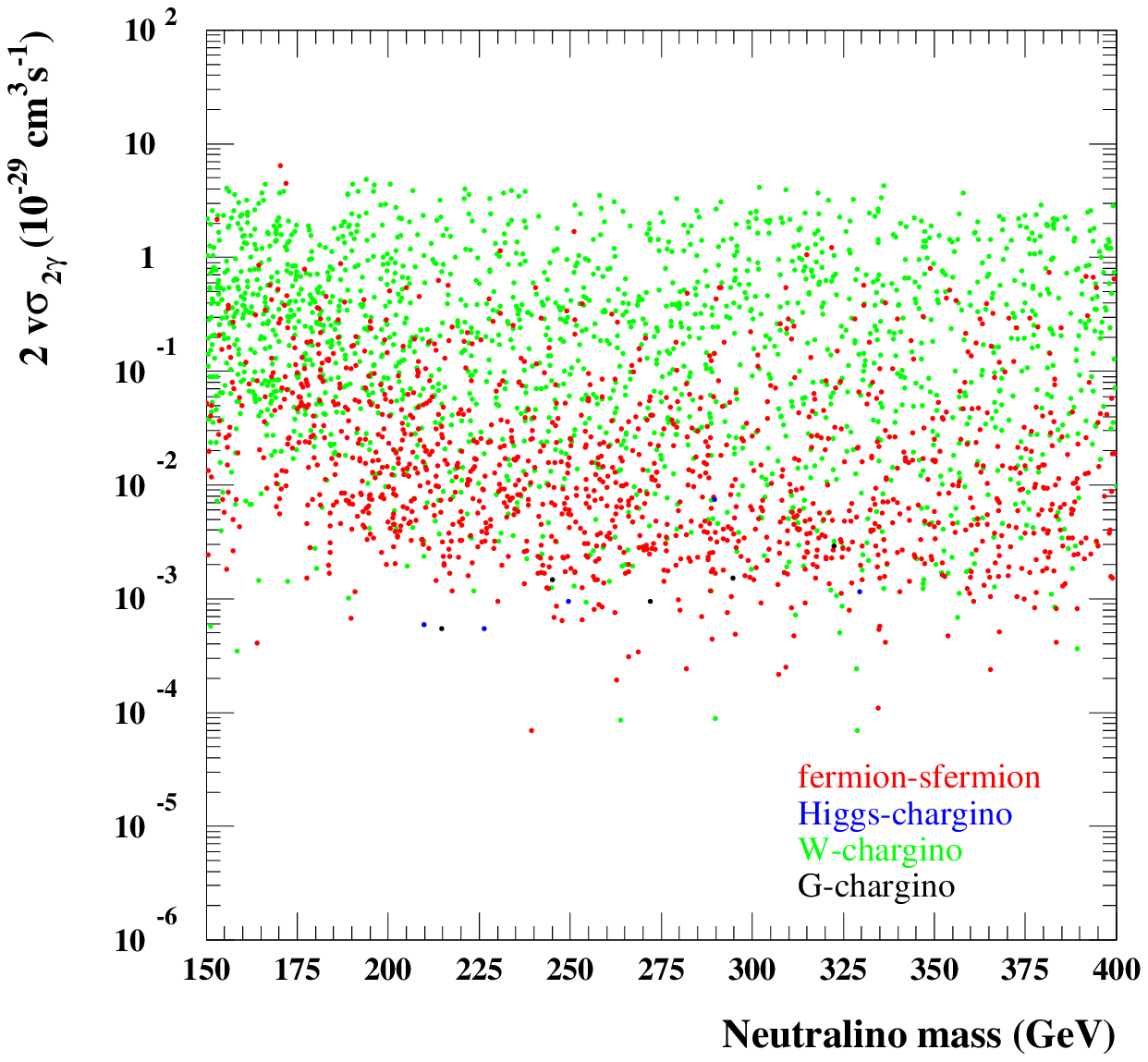,width=7cm}} \quad
       {\epsfig{file=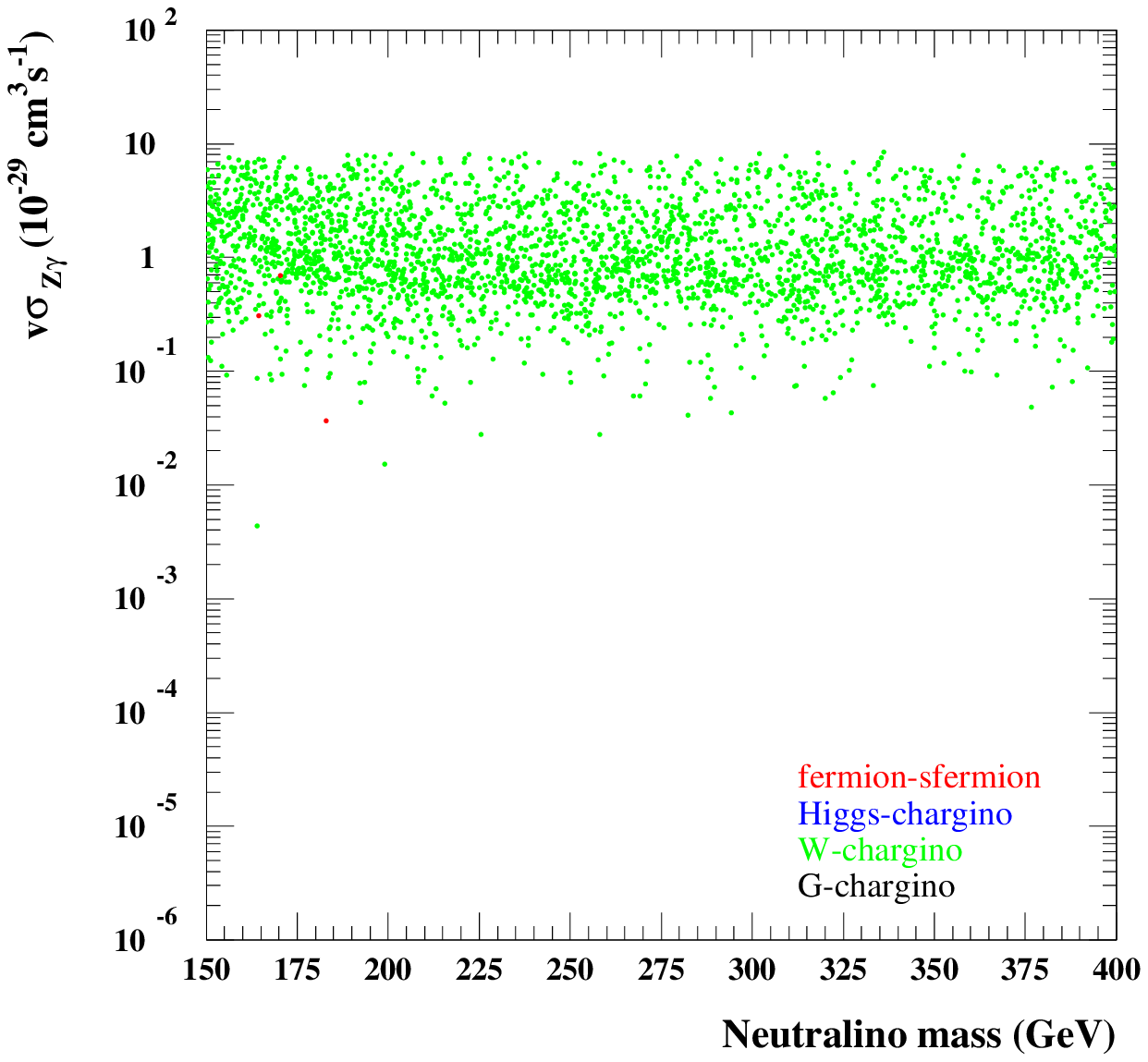,width=7cm}}}
\caption[]{$2\,v \sigma_{2\gamma}$ and $v \sigma_{Z\gamma}$ for a sample 
of mixed neutralinos in the mass range 150 - 400 GeV. For each model 
the class of diagrams which gives the main contribution to the cross 
section is indicated.} 
\label{fig:xsmixed}
\end{figure}

Whereas for higgsino-like neutralinos $2\,v \sigma_{2\gamma}$ is in 
most cases greater than $v \sigma_{Z\gamma}$, it is generally the 
opposite for mixed neutralinos. For this class of neutralinos in 
the mass interval $100 - 500$ GeV the branching ratios into the 
$Z\gamma$ final state is in most cases dominant over twice the 
branching ratio into $2\gamma$. Another interesting feature is 
that over the whole mass scale the value of $v \sigma_{Z\gamma}$ 
for mixed neutralinos is concentrated in a band not wider than two 
orders of magnitude, while the values of $v \sigma_{2\gamma}$ are 
more homogeneously distributed over 5 to 6 decades. This is visible 
also in Fig.~\ref{fig:zgx2gx} where the mixed neutralino band starts 
at about $6-8 \cdot 10^{-29}\;\rm{cm}^3 \rm{s}^{-1}$ for both 
$2\,v \sigma_{2\gamma}$ and $v \sigma_{Z\gamma}$ and extends down 
to to a value of $2\,v \sigma_{2\gamma}$ of 
$10^{-34}\rm{cm}^3 \rm{s}^{-1}$ for models that can give a  
$v \sigma_{Z\gamma}$ as high as $10^{-29}\;\rm{cm}^3 \rm{s}^{-1}$. 
We illustrate this point in Fig.~\ref{fig:xsmixed} where we consider 
a scan of models with mixed neutralinos of masses in the range 
$150 - 400$ GeV. In these figures we have indicated which is the class 
of diagrams that gives the main contribution to the cross sections 
(all the diagrams involved in the computation are shown in 
Refs.\,~\cite{lp} and~\cite{lp2}). In the case of $v \sigma_{Z\gamma}$ 
almost always the W boson-chargino loop diagrams are dominant over 
the other diagrams. For $2\,v \sigma_{2\gamma}$ instead, the 
contributions for the fermion-sfermion loop diagrams can be more 
frequently of the same order of magnitude as the W boson-chargino 
diagrams and severe cancellations can take place, with usually a 
small cross section when the fermion-sfermion diagrams are dominant. 
 From Fig.~\ref{fig:xsmixed} it is also evident that Higgs-chargino loop 
diagrams and Goldstone boson-chargino loop diagrams do not play a 
main role and this is true for all classes of neutralinos over the 
whole mass range.

Given the large size of our sample of scanned models, we expect the 
narrowness of the $Z\gamma$ band for large neutralino masses to be a 
real effect which could increase the discovery potential of present
and future ACTs. In the heavy mass range, it may be appropriate to sum 
the $\gamma\gamma$ and $Z\gamma$ contributions, since to resolve the 
two lines an energy resolution better than around 1 \% would be 
needed. 

Gaugino-like neutralinos give cross sections that rarely exceed, for 
both processes, a few times $10^{-30}\;\rm{cm}^3 \rm{s}^{-1}$. In this 
case the cross sections are poorly correlated and  the large 
spread in Fig.~\ref{fig:zgx2gx} is due mainly to 
gaugino-like neutralinos.

A final comment can be made about low mass neutralinos. 
Here $v \sigma_{Z\gamma}$ goes to zero at the threshold 
$M_{\chi}\to M_Z/2$ as expected because this limit corresponds 
to having no photon in the final state and the annihilation of 
neutralino pairs into a Z boson is forbidden by C parity (for 
clarity we have excluded models with neutralino masses lower than 
70 GeV from Fig.~\ref{fig:zgx2gx}). 
In addition, $2\,v \sigma_{2\gamma}$ has low 
values as well because the contribution from W boson-chargino loop 
diagrams becomes small (there is no contribution from the imaginary
 part and the real part is generally large only for heavy pure 
higgsino-like neutralinos).


\section{Milky Way halo models}\label{sec:halo}
The photon flux from neutralino annihilation in the galactic halo critically 
depends on the neutralino distribution. Since the flux per unit volume is 
proportional to the neutralino density squared, any enhancement of the 
density would result in a sharp increase in the photon flux; such an 
enhancement is provided by dark matter radial halo profiles that are peaked 
towards the galactic center. In this section we will examine in detail 
this possibility and compute the full dependence of the flux on the 
galactic coordinates in some selected cases.

The mass distribution in the Milky Way and the relative importance 
of \mbox{its} three components, the bulge, the disk and the halo, are poorly 
constrained by available observational data. Although the dynamics of 
the satellites of the galaxy clearly indicates the presence of a 
non-luminous matter component, a discrimination among the different radial 
dark matter halo profiles proposed in the literature is not possible at 
the time being ~\cite{binney}. Our approach is to assume that dark matter 
profiles are of a universal functional form and to infer 
the Milky Way dark matter distribution from the results of 
N-body simulations 
of hierarchical clustering in cold dark matter cosmologies. The predicted 
profiles in these scenarios have been tested to a sample of dark matter 
dominated dwarf and low-surface brightness galaxies which provide the best 
opportunities to test the spatial distribution of dark matter. Actually 
this field of research is in rapid evolution and slightly discrepant 
results have recently been presented \cite{carlberg,navarro,kravtsov,moore}; 
we will concentrate on 
the Kravtsov et al. profile \cite{kravtsov} and the Navarro et al 
profile \cite{navarro}, comparing the results with the canonical modified 
isothermal sphere profile which is widely used in dark matter calculations.

Among the general family of dark matter halo profiles
\beq
\rho(r) \propto \frac{1}{(r/a)^{\gamma}\;[1+(r/a)^{\alpha}]^
{(\beta-\gamma)/\alpha}},
\eeq
Kravtsov et al. \cite{kravtsov} propose a coreless profile with a mild 
singularity towards the galactic center with $\gamma \sim 0.2 - 0.4$; 
taking into account these two extremes, we will refer to the Ka profile 
as to the one which is defined by $(\alpha,\beta,\gamma)=(2,3,0.2)$, 
while the Kb profile is chosen as $(\alpha,\beta,\gamma)=(2,3,0.4)$. In 
previous work~\cite{navarro}, Navarro, Frenk and White had obtained a 
cuspy profile which has $(\alpha,\beta,\gamma)=(1,3,1)$ (hereafter NFW 
profile). Both results are not in agreement with the modified isothermal 
distribution, $(\alpha,\beta,\gamma)=(2,2,0)$  (hereafter Sp profile), 
extensively used  in the analyses of observed rotation curves, which 
is non singular with a core of radius $\sim$ a.

There are models \cite{bere} which have a more singular behavior near 
the galactic center, and which would give enormously enhanced rate in 
that direction. However, there is observational evidence against such 
singularities from cluster gravitational lensing and the rotation curves 
of dwarf spiral galaxies \cite{flores}. On the other hand, the discrepancy 
between the 1/r central cusp of the NFW profile and the experimental data 
from the dwarf spheroidal DDO~154 has been explained in Ref. \cite{bs} 
assuming an additional component of dark baryons. Of course, also a 
moderately steeper
profile like $1/r^{1.3}$, which is found in some simulations \cite{makino}
and which
has some theoretical underpinning \cite{evans}, may be acceptable. That 
would give higher estimates than the one we use for the neutralino-induced
fluxes, but we choose to be more conservative.

We will consider only the case of spherical profiles; introducing a 
flattening parameter may enhance the value of the flux but the effect 
is not expected to be dramatic for this neutralino detection method and we 
prefer not to introduce another factor of uncertainty.

The normalization constant of the halo profile, which we choose as the 
value of the halo density $\rho_0$ at our galactocentric distance $R_0$, 
the core radius $a$ and $R_0$ itself are the three parameters which will be 
relevant to give a prediction for the gamma line flux. N-body simulations 
indicate that there may be a correlation between normalization constant 
and core radius, but we are not in the position of extracting a firm 
numerical prediction for the Milky Way. After choosing the functional 
form of the halo profile, we determine the regions in the parameter space 
which are allowed by the assumptions that follow below.

We follow Dehnen and Binney \cite{binney} and assume that the total mass
 of the galaxy inside 100 kpc is $7 \pm 2.5 \times 10^{11} {\rm M_{\odot}}$.
 This estimate is obtained by combining the result of 
Kochanek~\cite{kochanek} who used the observational constraints from 
the velocity distribution of the Milky Way's satellites, the local 
escape velocity of stars and the Local Group timing model, and the 
result inferred in Ref.~\cite{lin} 
from the dynamics of the Magellanic Clouds and Stream. Both estimates 
are to some extent model dependent and this is the reason for the large 
error bound chosen. We then suppose that, for each matter distribution 
model, on average ten per cent of the total mass inside 100 kpc is due to 
the contributions of the disk and the bulge (value obtained from Table 4 
in \cite{binney}) and derive a value for the the mass of dark matter halo
 inside 100 kpc
\beq
M_h (r<100\, \rm{kpc}) = (6.3 \pm 2.5) \times 10^{11} {\rm M}_{\odot} \label{eq:mass}
\eeq

A second constraint is given in terms of the contribution of the halo to
 the local rotation curves which can be determined in terms of the observed 
value of the local rotation velocity $\Theta_0$ and from the measurements
 of the contribution of the disk (which is dominant over the contribution 
from the bulge):
\beq
v_h^2 (R_0) = \Theta_0^2 - v_{d}^2 (R_0) 
\eeq
The IAU standard values for $\Theta_0$ and our galactocentric distance 
${\rm R}_0$, based on a review by Kerr and Lynden-Bell \cite{kerr} in 1986, 
are respectively $220 \pm 20\,\rm{km}\, \rm{s}^{-1}$ and $8.5 \pm 
1.1\,{\rm kpc}$. The two quantities are not independent variables, their 
ratio is fixed by local stellar kinematic measurements of the Oort 
constant. 
In more recent estimates lower values have been obtained: Reid~\cite{reid} 
has found ${\rm R}_0 = 7.2 \pm 0.7\,{\rm kpc}$, while in a very recent 
work Olling and Merrifield~\cite{oll} has found ${\rm R}_0 = 7.1 
\pm 0.4\,{\rm kpc}$ and $\Theta_0 = 184 \pm 8\,\rm{km}\, \rm{s}^{-1}$. 
Even the value of the contribution of the disk to the local rotation 
velocity has large uncertainties and is to some extent model dependent. 
The value we can infer from Kuijken and Gilmore \cite{kui}, and 
Gould \cite{gould}, who examine a data set of velocities of local stars 
as a function of their height over the galactic plane, is in the range 
$v_d(R_0) = 137\pm 16\,\rm{km}\, \rm{s}^{-1}$. 
Maximum disk models can give a 
higher $v_d(R_0)$, while in thin disk models the value can be lower. We 
will accept the quoted value without considering disk models in detail.
Combining the values of $\Theta_0$ and $v_d(R_0)$ we obtain the following 
ranges of values of $v_h(R_0)$ for a given galactocentric distance ${\rm R}_0$:
\beqa
v_h(R_0) \sim 128 - 207\, \rm{km}\, \rm{s}^{-1} &\;\mbox{\normalsize if}\;\;\; R_0 = 8.5\, \rm{kpc} \label{eq:vel85}
\eeqa
\beqa
v_h(R_0)\;\sim\; 86 - 149\, \rm{km}\, \rm{s}^{-1} & \;\mbox{\normalsize if}\;\;\; R_0 = 7.1\, \rm{kpc} \label{eq:vel71}
\eeqa

\begin{figure}[htb]
 \epsfig{figure=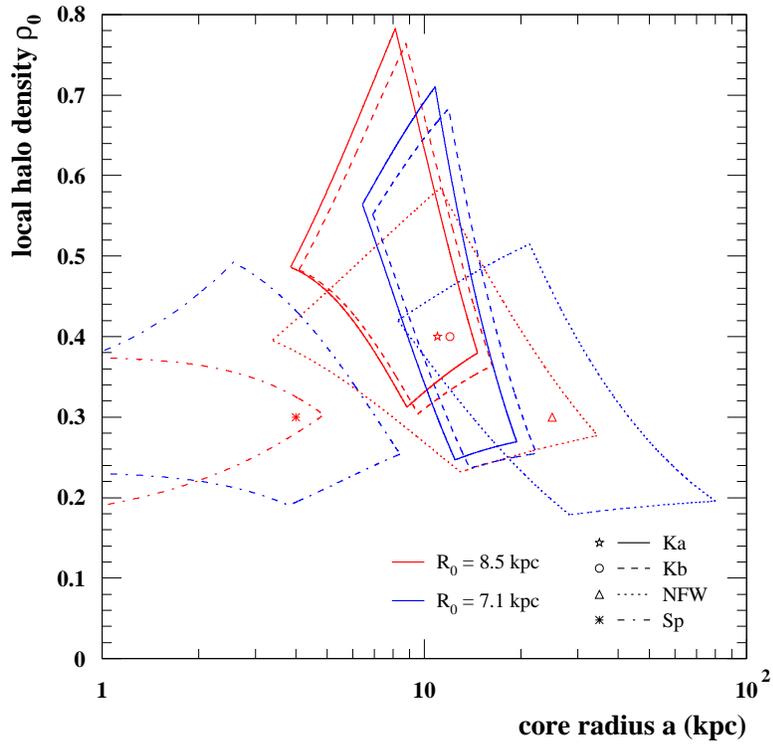,width=12cm}
\caption[]{Allowed values of the parameters $\rho_0$, the local halo density, 
and $a$, the core radius, for the four halo profiles considered in 
Section~\ref{sec:halo}. The allowed regions vary as a function of the 
third parameter which enters the discussion, the galactocentric distance 
of the solar system $R_0$; we plot the regions corresponding to 
$R_0 = 8.5\, \rm{kpc}$ which extend to lower values of $a$ and 
to $R_0 = 7.1\, \rm{kpc}$ which allow higher values of $a$. The markers indicate the halo profiles which are considered in Fig.~\ref{fig:jpsi}.} 
\label{fig:arho}
\end{figure}

In Fig.~\ref{fig:arho} we show, for each of the  halo models considered, 
the regions in the plane ($a$, $\rho_0$) which are compatible with the 
limits given in Eq.~(\ref{eq:mass}) and 
Eq.~(\ref{eq:vel85}),~(\ref{eq:vel71}).
A common feature of all the profiles is that the allowed regions are 
shifted to lower values of the core radius increasing the galactocentric 
distance from 7.1 kpc to 8.5 kpc. Focusing for instance on the Ka profile 
with $R_0 = 8.5\,\rm{kpc}$, the values of $a$ and $\rho_0$ which are 
compatible with our constraints fall in an area with a diamond-like shape, 
where the upper left bound and the lower right bound are given 
respectively by the maximum and minimum values in Eq.~(\ref{eq:vel85}) , 
while the upper right and lower left are due to Eq.~(\ref{eq:mass}). A 
similar behavior is seen in all the other cases. For the Sp profile a core 
radius equal to zero is not excluded, but this is an extremely unrealistic 
case and we have decided to arbitrarily fix a lower value of 
$a = 1\,\rm{kpc}$.

\begin{figure}[htb]
 \epsfig{figure=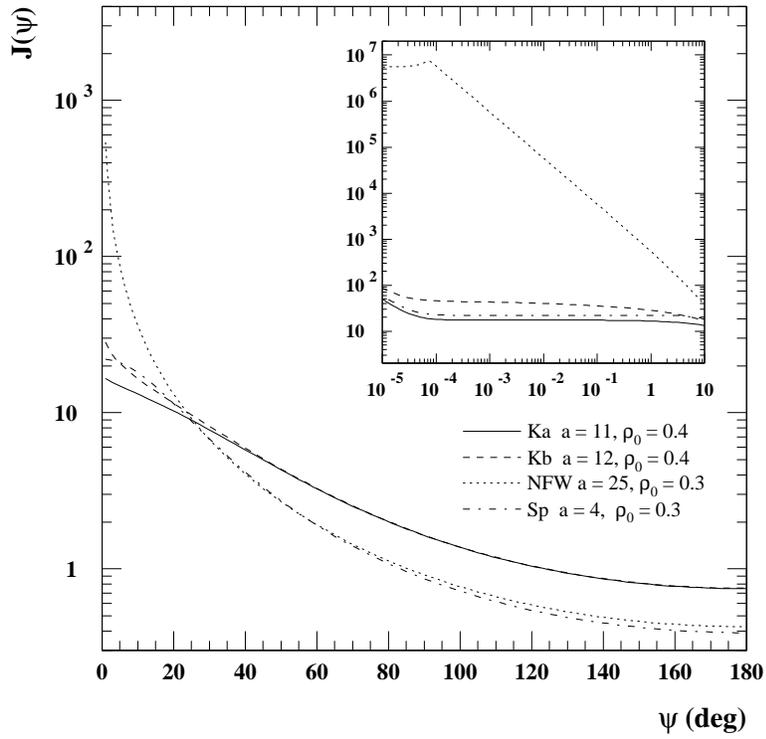,width=12cm}
\caption[]{Function $J(\psi)$ as defined in Eq.~(13) for the four halo 
profiles considered in Section~\ref{sec:halo}. The choice of the parameters
$a$ and $\rho_0$ is indicated in the figure, while we have fixed $R_0 = 8.5$ kpc.}
\label{fig:jpsi}
\end{figure}

The gamma ray flux from neutralino annihilation in the galactic halo is 
given by:
\beq
\Phi_{\gamma}(\psi) = \frac{N_{\gamma}\;v\sigma}{4\pi M_\chi^2} 
\int_{line\;of\;sight}\rho^2(l)\; d\,l(\psi)
\eeq
where $\psi$ is the angle between the direction of the galactic center and 
that of observation, and where $N_{\gamma} = 2$ for 
$\chi\,\chi\rightarrow \gamma\,\gamma$, $N_{\gamma} = 1$ for 
$\chi\,\chi\rightarrow Z\,\gamma$. 
Separating the dependence on the halo model from the part which is related
 to the the values of the cross section and the neutralino mass, we 
rewrite it as
\beqa
\Phi_{\gamma}(\psi) & \simeq & 1.87 \cdot 10^{-11}\left( \frac{N_{\gamma}\;v\sigma}
{10^{-29}\ {\rm cm}^3 {\rm s}^{-1}}\right)\left( \frac{10\,\rm{GeV}}
{M_\chi}\right)^2 \cdot \nonumber \\
&&\;\;\;\;\;\; {\cdot}\, J\left(\psi\right)\;\rm{cm}^{-2}\;\rm{s}^{-1}\;\rm{sr}^{-1}
\eeqa
where we have defined the dimensionless function 
\beq
J\left(\psi\right) = \frac{1} {8.5\, \rm{kpc}} 
\cdot \left(\frac{1}{0.3\,{\rm GeV}/{\rm cm}^3}\right)^2
\int_{line\;of\;sight}\rho^2(l)\; d\,l(\psi)\;.
\label{eq:jpsi}
\eeq
Picking in Fig.~\ref{fig:arho}, for the four halo models, a couple of 
values ($a$, $\rho_0$) which gives rather conservative results and 
choosing $R_0 = 8.5\,\rm{kpc}$, we show in Fig.~\ref{fig:jpsi} in each 
case the angular dependence of the function $J$. The profiles have been 
considered valid up to the capture radius of the black hole at the 
galactic center which is about $0.01\, {\rm pc}$ for a mass of Sgr A$^*$ 
${\rm M} \simeq 10^6\, {\rm M}_{\odot}$. Modifications to 
Eq.~(\ref{eq:jpsi}) due presence of the black hole are in detail 
in the subsection below.

\begin{figure}[htb]
 \epsfig{figure=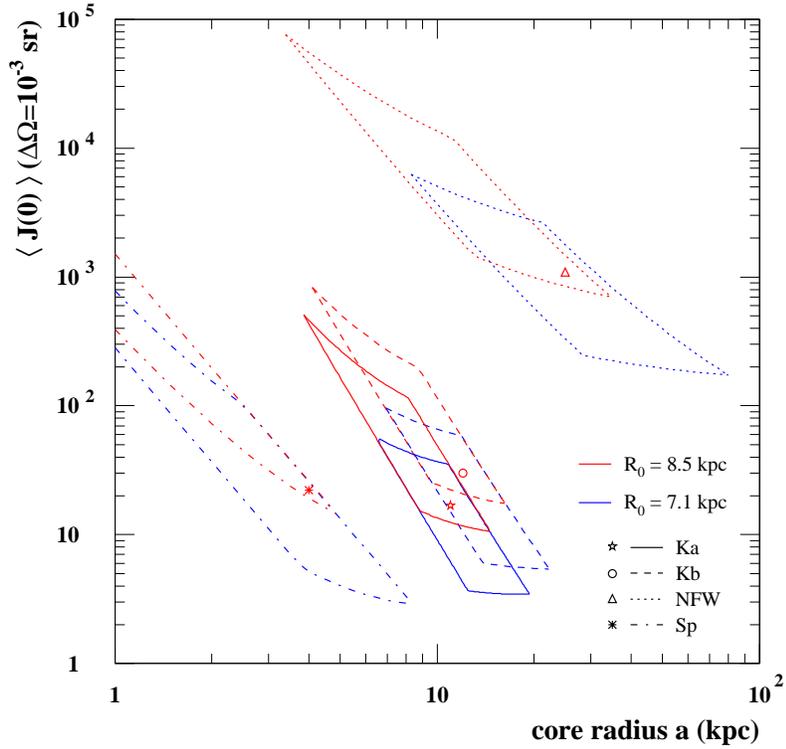,width=12cm}
\caption[]{$\langle\,J\left(0\right)\,\rangle\,$ as defined in Eq.~(14)  
for $\Delta\Omega = 10^{-3} sr$ and as a function of the core radius $a$ 
for the four halo profiles considered in Section~\ref{sec:halo}. Each contour is given by the maximal and minimal values of the parameter $\rho_0$ as given in Fig.~\ref{fig:arho}, for the two choices of $R_0$.}
\label{fig:jd3}
\end{figure}

\begin{figure}[htb]
 \epsfig{figure=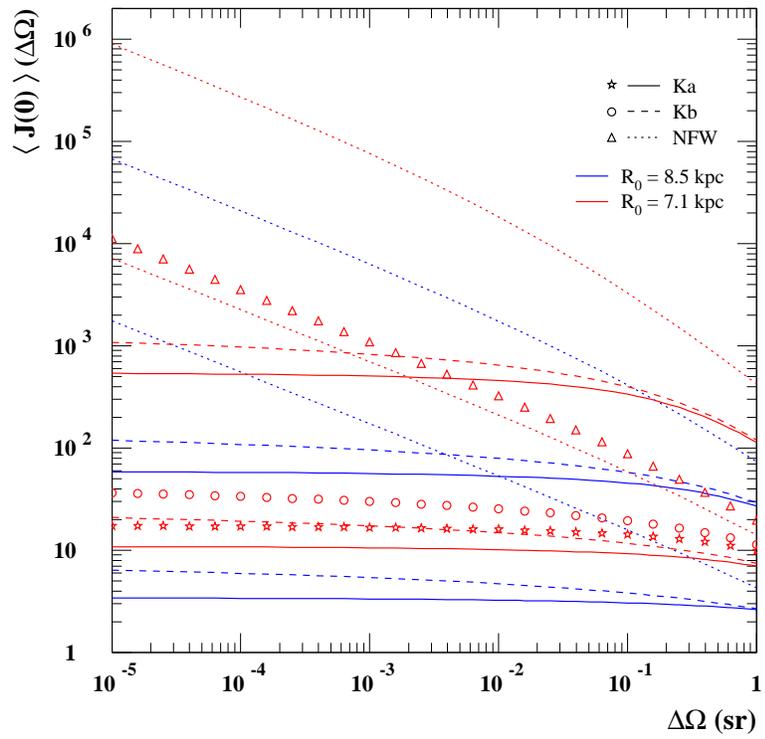,width=12cm}
\caption[]{$\langle\,J\left(0\right)\,\rangle\,$ as a function of 
$\Delta\Omega$ for the halo profiles defined by the maximal and 
minimal values of the core radius allowed in Fig.~\ref{fig:arho} and for those 
considered in Fig.~\ref{fig:jpsi}. 
Curves for both choices of $R_0$ are shown.}
\label{fig:jd}
\end{figure}

The maximum flux will be clearly in the direction of the galactic center. 
The relevant quantity for a measurement is, rather than $J\left(0\right)$, 
the integral of $J\left(\psi\right)$ over the solid angle given by the 
angular acceptance of a detector which is pointing towards the galactic 
center. We consider therefore the function
\beq
\langle\,J\left(0\right)\,\rangle\,(\Delta\Omega) = \frac{1} {\Delta\Omega} 
\;\int_{\Delta\Omega}\,J\left(\psi\right)\,\; d\Omega
\eeq
where $\Delta\Omega$ is the angular acceptance of the detector.
Fixing for an ACT detector a typical value of $\Delta\Omega = 10^{-3}$, 
in Fig.~\ref{fig:jd3} we plot $\langle\,J\left(0\right)\,\rangle$ versus 
the core radius $a$ for the allowed regions in Fig.~\ref{fig:arho} and 
for both choices of the galactocentric distance. As can be seen the NFW 
profile can give very high values of $\langle\,J\left(0\right)\,\rangle$ 
compared to the other halo models. In Fig.~\ref{fig:jd} we plot 
instead $\langle\,J\left(0\right)\,\rangle$ as a function of $\Delta\Omega$
 for the conservative halo models shown in Fig.~\ref{fig:jpsi} together 
with the models corresponding to the maximum and minimum core radius 
allowed in Fig.~\ref{fig:arho} for each profile.

\subsection{Effects of a central black hole}

A small complication occurs when we try to compute the line-of-sight
integral directly towards the galactic center.
There are strong indications that a black hole of mass 
$\sim 10^6\ \rm{M}_\odot$
resides near the center of the Milky Way. The interplay between this black
hole and the dark matter halo is difficult to model - certainly it is 
beyond the scope of this paper. One effect of the black hole, however, is
to gravitationally lens $\gamma$ rays that originate from behind the 
galactic center, in a cone of  angular size on the order of that of 
the Einstein ring. This is given by (see, e.g., \cite{schneider})
\beq
\theta_E=\sqrt{{4GM\over c^2}{D_{ds}\over D_dD_s}},
\eeq
where $G$ is Newton's constant, $M$ the mass of the black hole,
$D_{ds}$, $D_d$, $D_s$ are the lens-to-source, lens and source distances,
respectively.

In general, for a point source, a double image is produced which usually 
cannot be resolved (the situation is similar to that of microlensing
of \mbox{MACHOs} in the halo). The resulting magnification may, however, be
detectable and is given by
\beq
\mu={u^2+2\over u\,\sqrt{u^2+4}},
\eeq
where $u=\beta/\theta_E$, with $\beta$ the undeflected angle to the source.
Since gravitational lensing is a purely geometrical effect, it is of 
course also operative for $\gamma$ rays.
Returning to our line-of-sight integral Eq.~(\ref{eq:jpsi}), we can now
for angles $\psi$ less than the Einstein angle 
$\theta_E(l)=\theta_0\sqrt{(l-R_0)/l}$ (with $\theta_0$ around 1 arcsec
for $M=10^6\,\rm{M}_\odot$ and $R_0=8.5$ kpc)
split the integral into two parts, from $0$ to the distance of the galactic center $R_0$ and from $R_0$ to the end of the halo. The second part contains the 
amplification factor $\mu(l)$, thus for small angles we replace
$J(\psi)$ by $K(\psi)$ where
\beqa
K(\psi)&=&\frac{1} {8.5\, \rm{kpc}} 
\cdot \left(\frac{1}{0.3\,{\rm GeV}/{\rm cm}^3}\right)^2 \cdot
\Bigg[\,\int_0^{R_0}\rho^2(l)\;dl(\psi)\;+ \nonumber\\
&&\;\;\;+\;\int_{R_0}^\infty\rho^2(l)\;\left[ 
{\left({\psi\over\theta_E(l)}\right)^2+
2\over\left({\psi\over\theta_E(l)}\right)\sqrt{\left({\psi\over\theta_E(l)}\right)^2+4}}\;\right] dl(\psi)\Bigg]\;.
\eeqa
In Fig.~\ref{fig:jpsi} this expression has been used for the 
smallest angles
displayed. As can be seen, it causes a mild enhancement for the most 
non-singular halo models and cannot compensate the turnoff of the rise
of the $NFW$ profile due to the black hole itself.


\section{Present and Planned ACT Detectors}\label{sec:planned}

The present high energy gamma-ray experiments, the {\it Energetic Gamma-Ray
Experiment Telescope} \,(EGRET) and the Whipple 10m atmospheric Cherenkov
Telescope (e.g.~\cite{Cawley}),
 lack the sensitivity to detect the annihilation lines
fluxes predicted for most of the allowed supersymmetric models
and halo profiles.
Our aim is to show that the order of magnitude improvement in the flux
sensitivity and the substantial improvements in angular resolution
 in the next generation of both ground based and satellite 
gamma-ray experiments, will allow to explore large portions of
the \mbox{MSSM} parameter space searching for the annihilation-line flux,
if a halo model which provides an increase of the dark matter density 
towards the galactic center is considered.

In this section we focus on Atmospheric Cherenkov
Telescopes (ACTs), such as  an upgraded version of the Whipple 
detector currently
under construction (GRANITE-III) as well as
proposed arrays of ground-based atmospheric Cherenkov detectors
VERITAS~\cite{Weekes} 
and HESS~\cite{Aharonian}.
We will consider in Section~\ref{sec:glast} 
the potentiality of the proposed {\it Gamma-Ray Large Area Space Telescope}
\,(GLAST) experiment.
Ground based and satellite experiments are to some 
extent complementary for our purpose.
While satellite experiments offer a lower energy threshold, superior energy
resolution, and the possibility of relatively long 
exposure times, the relatively
small effective area $\sim 1$~m$^2$ limits the sensitivity 
at high energies. On the other hand ACTs offer the 
potential for very large effective areas
$\sim$0.3~km$^2$, but at a higher energy threshold 
($E_\gamma >$~250~GeV). 

We summarise in Table~\ref{tab:act} the approximate characteristics for existing and proposed ACT gamma-ray observatories.

\begin{table*}
\caption[]{Characteristics of Atmosferic Cherenkov Telescopes$^\dagger$}
\begin{center}
{\footnotesize
\begin {tabular}{llll}
\hline
\hline
    & Whipple & GRANITE-III & VERITAS or HESS \\
\hline
Effective Area (cm$^2$) & 3.5$\times 10^8$ & 5$\times 10^8$ & 7$\times 10^8$ \\
Energy Resolution ($\sigma_E/E$) &  30\% & 20\%$^a$ & 15\%$^b$ \\
Angular Resolution ($\sigma_\theta$) &  0.14 & 0.1$^c$ & 0.07$^d$ \\
Energy Threshold & 250~GeV & 150~GeV & 50~GeV \\
Field of View (sr) & 0.001 & 0.004 & 0.004 \\
\hline
\multicolumn{4}{l}{$^a$ Estimated from Ref.~\cite{goret}.} \\
\multicolumn{4}{l}{$^b$ Resolution at 100~GeV~\cite{Aharonian}.} \\
\multicolumn{4}{l}{$^c$ Estimated from CAT
 observations of the CRAB nebula~\cite{Barrau}.} \\
\multicolumn{4}{l}{$^d$ Angular resolution at 
300 GeV from simulations
of a 4 telescope array \cite{Aharonian}.} \\
\multicolumn{4}{l}{$^\dagger$ Average quantities estimated for the
energy ranges: $E_\gamma >$300~GeV for Whipple;} \\ 
\multicolumn{4}{l}{$ E_\gamma > 150$~GeV for GRANITE-III; 
$E_\gamma > 75$~GeV for VERITAS or HESS} \\
\end{tabular}
}
\end{center}
\label{tab:act}
\end{table*}

For the observation of the galactic centre 
which is treated as
a point source, we compute the minimum detectable flux
$N_\gamma(E)$ using the prescription that,
for an exposure of $t$ seconds made with an instrument of
effective area $A_{\rm eff}$ and angular resolution
$\sigma_\theta$ (corresponding to a 68\% acceptance of events)
with corresponding angular acceptance $\Delta\Omega = \pi \sigma_\theta^2$, 
the conditions for detection are that the significance of the detection
exceed 5$\sigma$ and that the number of detected photons exceeds 25 events:
\begin{equation}
{\rm Significance\ } =
{N_\gamma A_{\rm eff} t \times 0.68^2
\over (A_{\rm eff} t)^{1/2} \left(dN_{\rm bg}/d\Omega \times \Delta\Omega
\right)^{1/2}} \ge 5
\label{eq:sig}
\end{equation}
\begin{equation}
N_\gamma A_{\rm eff} t  \ge  25. 
\label{eq:min}
\end{equation}
Here, $dN_{\rm bg}/d\Omega$ is the integral number of
background
events (per unit solid angle) falling under the annihilation line.
If the dominant source of background has a differential spectrum
${d^2N_{\rm bg}/ dE d\Omega} = N_0 E^{-\delta}$,
and if the energy resolution of the instrument is $\sigma_E/E$,
then the background under a line at energy $E_0$ (i.e., in the interval
$[E_0-\sigma_E, E_0+\sigma_E]$ containing 68\% of the signal)
is given by
\begin{equation}
{dN_{\rm bg}\over d\Omega} = {N_0\over (\delta-1)}E_0^{-\delta+1}\times
\eta(\sigma_E/E,\delta) \quad ,
\end{equation}
where
\begin{equation}
\label{eq:resolution}
\eta(\sigma_E/E,\delta)=
\left[(1-\sigma_E/E)^{-\delta+1} - (1+\sigma_E/E)^{-\delta+1}\right]
\end{equation}
gives the
the background reduction relative to the integral
background spectrum (e.g., $\eta = 0.35$ for an energy resolution of 10\%).

Atmospheric Cherenkov detectors employ large ($\sim$10~m) optical reflectors
to image flashes of Cherenkov light from electromagnetic showers formed as high
energy $\gamma$-rays interact in the earth's atmosphere.  The Cherenkov light
pool from a $\gamma$-ray (at normal incidence) covers an area of roughly
5$\times10^8$~cm$^2$ on the ground, and defines the sensitive area of
the detector.  By making use of the
distinctive differences in the shower images, a
$\gamma$-ray signal can be extracted from the large isotropic background of
hadronic showers (e.g.~\cite{Reynolds}).
The point of origin of each shower can also be uniquely reconstructed
from the orientation and shape of the shower image to a precision of
$\sigma\,\approx$ 0.14$^\circ$ for a single telescope (e.g.~\cite{Buckley}),
 or roughly a factor of
$\sqrt{N}$ better
when $N$ telescopes provide stereoscopic views of the same shower image.
The total level of detected Cherenkov light 
is roughly proportional to the energy of the primary
gamma-ray shower and provides an energy resolution typically of
$\sigma_E/E\approx$20--40\%.  A further improvement in this energy
resolution may
be realized with stereoscopic measurements which determine the height of
the maximum shower development.
Typically observations are
made in a differential mode where each observation on-source is followed
by an observation of equal duration displaced in right ascension
from the source direction, so that the control observations
sample the same range of azimuth and elevation angles.  The number
of candidate gamma-ray events is taken to be the difference between the
number of events passing the data-selection criteria
on-source and in the control observations.  Using this technique one
can determine the two-dimensional distribution of gamma-ray events on
the sky, and can reconstruct the energy spectrum of these events assuming
that the control region does not contain a significant contribution 
from diffuse gamma-rays or from a number of point sources.

The energy threshold of atmospheric Cherenkov detectors is determined
by the condition that the Cherenkov light signal must exceed the level
of fluctuations in the night sky background light.  Since both the
signal and the background light level are proportional to the mirror
area $A_{\rm m}$, the signal to noise ratio depends on 
$\sqrt{A_{\rm m}}$ and the energy threshold is inversely proportional
to this quantity.  Since the duration of the Cherenkov pulses is
very small (on the order
of a few nsec), an additional reduction in the energy threshold can be 
realized by decreasing the signal integration time.
For the largest imaging telescope (the Whipple 10m reflector) the
energy threshold is approximately 250~GeV.  The use of arrays of telescopes
operated in coincidence will result in a larger
effective mirror area which when combined with the use of faster electronics
and higher resolution cameras will result in a reduction in the energy
threshold to perhaps 50~GeV for an array of 10m telescope.

Arrays of telescopes also offer the potential for increasing the effective
area and improving the background rejection.  $N$ widely spaced telescopes
operated independently result in an
increase in the effective area by $N$.  As the telescopes are moved closer
together, the effective area is reduced but the hadronic rejection can
be dramatically improved through better characterization of the
shower development with stereoscopic imaging.  For the VERITAS (HESS)
arrays, 9 (16) 10~m telescopes will be used in a configuration where
each gamma-ray event within the sensitive area will result in good
shower images in four of the telescopes in the array.  In such a mode
of operation,
we estimate a total effective area of $7\times 10^8$~cm$^2$ and an additional
hadronic rejection by a factor of $\sim 4$ through improvements in angular
resolution and by another factor of $\sim 4$ through better characterization
of the shape of the gamma-ray shower (e.g. Ref.~\cite{Aharonian}).

The sensitivity of atmospheric Cherenkov detectors is determined by
a relatively large background of misidentified gamma-like hadronic showers 
and from cosmic ray electrons, which dominate at lower energies.
The contribution of the diffuse gamma-ray 
background is included in the sensitivity calculation, but 
has a relatively steep spectrum $\sim E^{-2.7}$ and does not 
contribute significantly to the background for atmospheric Cherenkov
telescopes.
Even if the diffuse spectrum of the inner galaxy is significantly harder than
the cosmic-ray spectrum~\cite{hunter}, this conclusion is unchanged.

From data taken with the Whipple 10~m telescope, the 
measured event rates (after rejection of background) and the effective angular
aperture and effective area (determined from Monte Carlo simulations) can
be combined to derive the background rate for gamma-like hadronic showers:
\begin{equation}
{dN_{\rm had}\over d\Omega}(E>E_0) = 6.1\times10^{-3}\epsilon_{\rm
had}\left({E_0\over 1\ {\rm GeV}}\right)^{-1.7}
 {\rm cm}^{-2}{\rm sec}^{-1}{\rm sr}^{-1}\quad .
\end{equation}
The factor of $\epsilon_{\rm had}$ is introduced 
to account for improved hadronic rejection in a future detector
compared with the current Whipple detector.  As the energy threshold
is reduced, the relative efficiency of hadronic showers for producing
Cherenkov light is reduced with respect to gamma-ray showers.  While
this effect would improve the sensitivity at low energies, the background
rejection from the image analysis generally deteriorates.  For
simplicity we have neglected these effects. 

The showers initiated by cosmic-ray electrons are
indistinguishable from gamma-rays, and these events can only be rejected on
the basis of their arrival directions. 
Cosmic-ray electrons have a steeper spectrum than
that of cosmic ray nuclei and become the dominant background at lower
energies.  From Ref.~\cite{longair} 
the integral spectrum for electrons is
\begin{equation}
{dN_{\rm e^-}\over d\Omega}(E>E_0) =
3.0\times10^{-2}\left({E_0\over
 1\ {\rm GeV}}\right)^{-2.3}
 {\rm cm}^{-2}{\rm sec}^{-1}{\rm sr}^{-1}\quad .
\end{equation}
(at $E=250\ {\rm GeV}$, the electron flux is a factor of 5 below the
hadronic background.)

\begin{figure}
 \centering
 \mbox{\subfigure{\epsfig{file=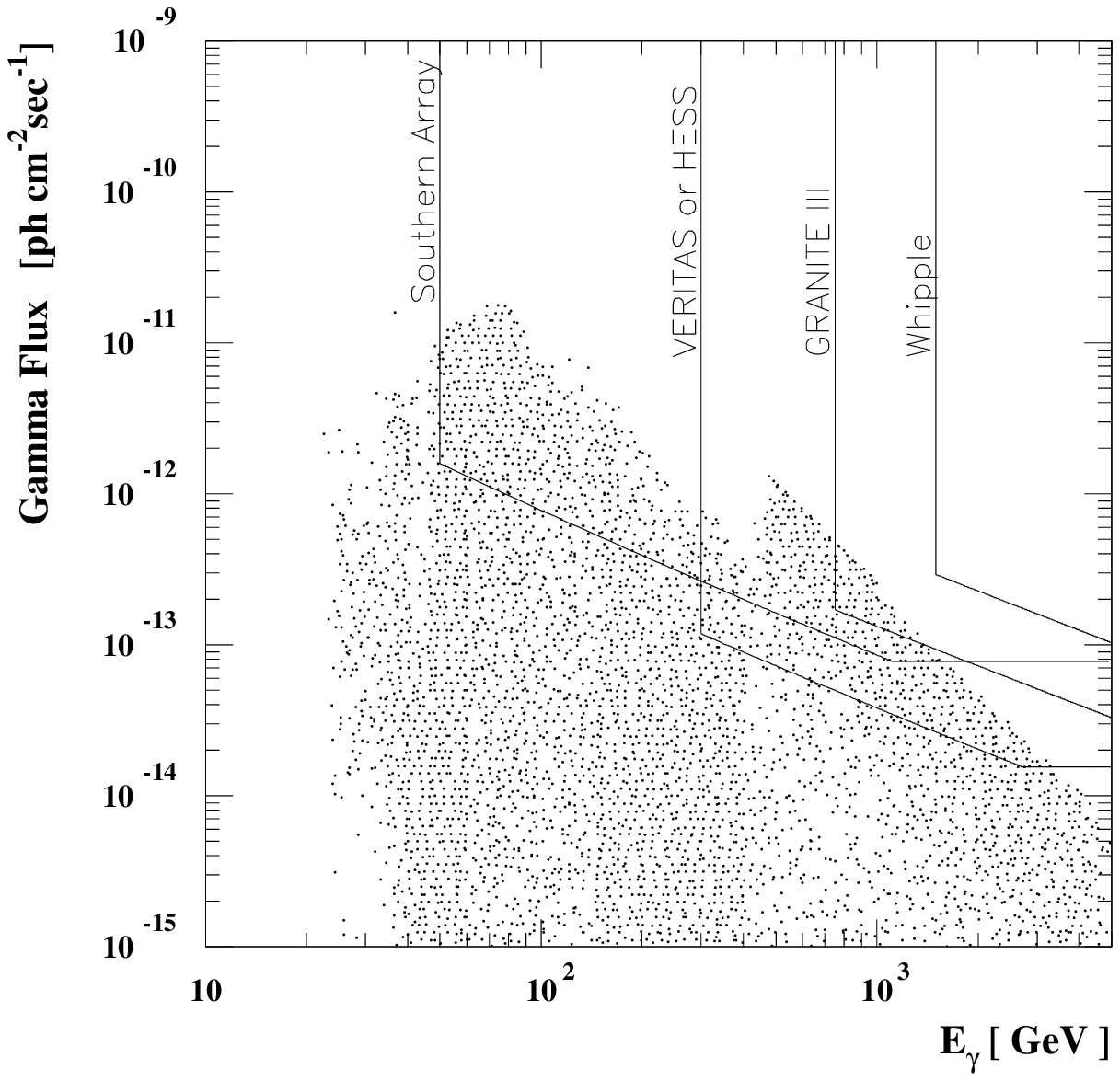,width=7cm}} \quad
       \subfigure{\epsfig{file=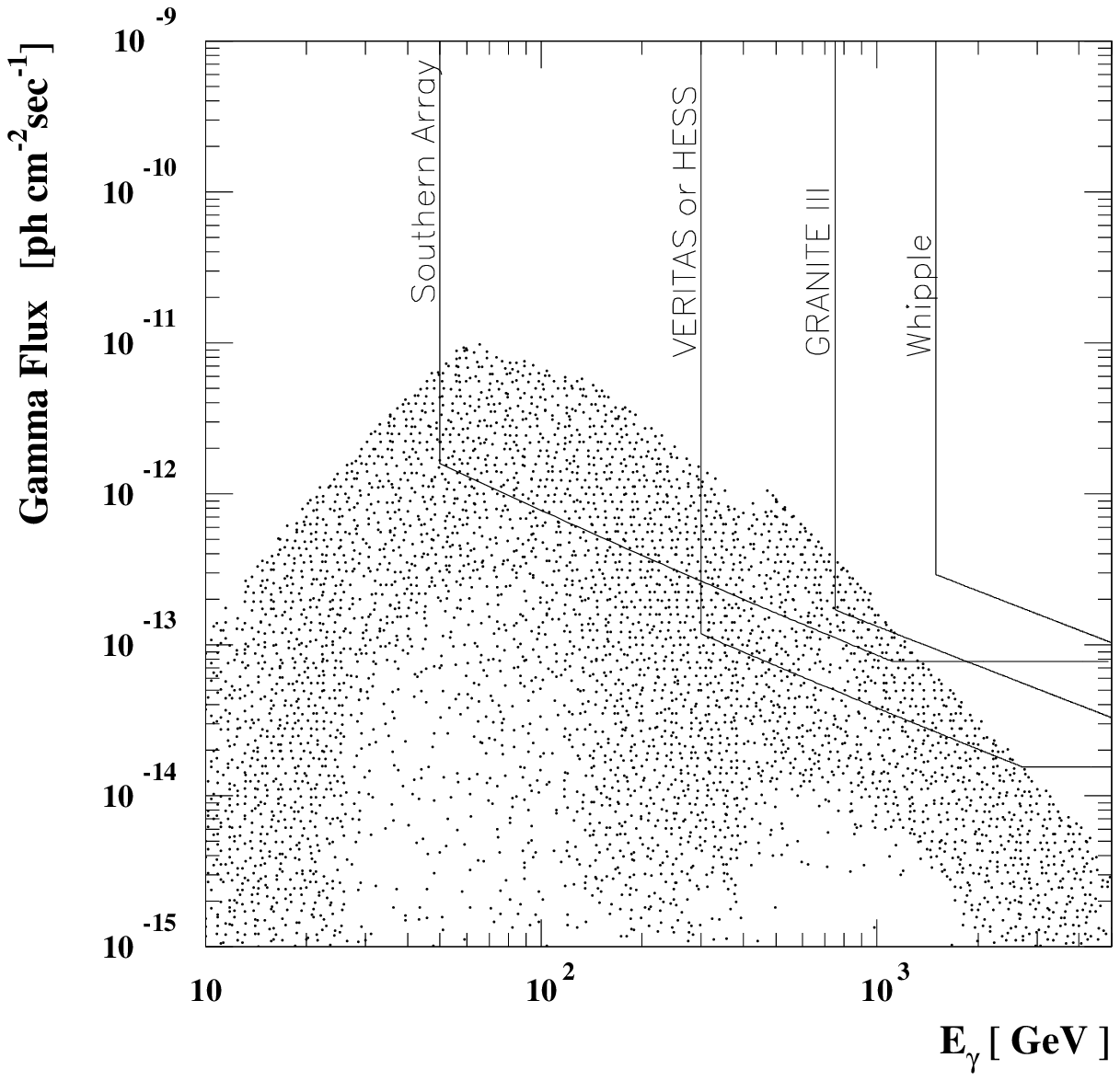,width=7cm}}}
\caption[]{Gamma ray flux from a $10^{-5}$ sr cone encompassing
 the galactic center for the $2\gamma$ (on the left) and the $Z\gamma$ 
annihilation line (on the right). The NFW halo profile giving the maximal flux 
has been assumed. The solid lines show the $5\sigma$ sensitivity curves 
of the ACT detectors described in the text.} 
\label{fig:act}
\end{figure}

\begin{figure}
 \epsfig{figure=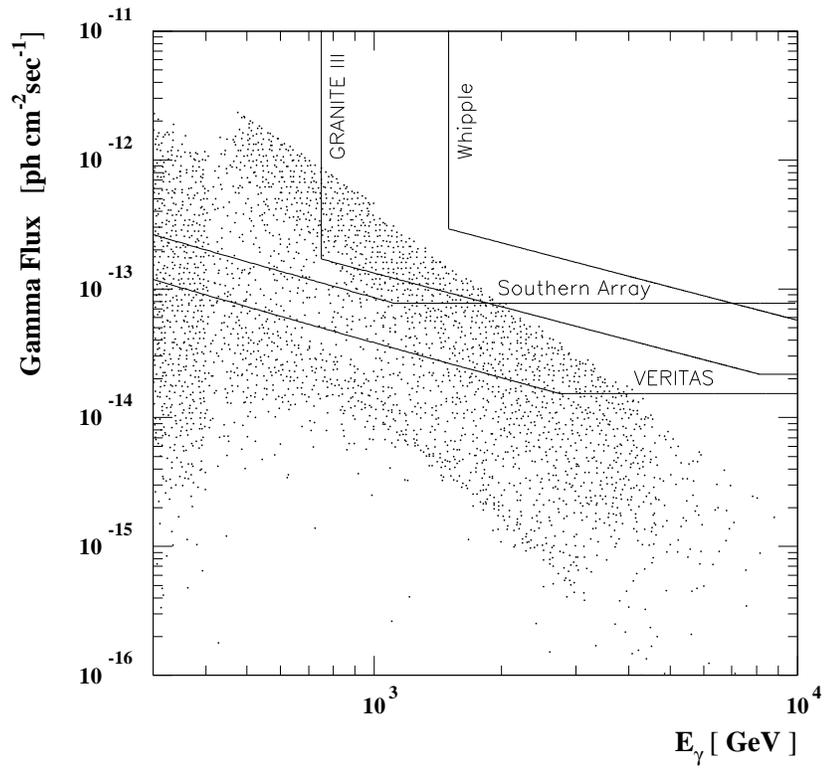,width=12cm}
\caption[]{Gamma ray flux from a $10^{-5}$ sr cone encompassing
 the galactic center summing the contributions of the $2\gamma$ and 
the $Z\gamma$ annihilation lines for heavy neutralinos.  
The NFW halo profile giving the maximal flux 
has been assumed.} 
\label{fig:acthm}
\end{figure}

Combining the hadron and electron background and taking into account 
the small contribution due to the diffuse gamma ray background,
the total background under the annihilation line at $E_0$, is
\beqa
\label{eq:linenoise}
\lefteqn{{dN_{\rm bg}\over d\Omega}\times \Delta\Omega  = 8.5\times 10^5\eta\left[ (\epsilon_{\rm had}+0.033)\left({E_0\over
100\ {\rm GeV}}\right)^{-1.7} +\,0.32\cdot \right.} \nonumber \\
&& \left. \cdot\left({E_0 \over 100\ {\rm GeV}}\right)^{-2.3}\right] 
\left({A_{\rm eff}\over
3.5\times 10^8 {\rm cm}^2}\right)\left({t\over 10^6 {\rm
sec}}\right)\left({\Delta\Omega\over 0.001 {\rm sr}}\right)
\eeqa
This can be used together with Eqs.~\ref{eq:sig} and \ref{eq:min}  
to derive the flux sensitivity curves of the Whipple atmospheric Cherenkov
detector as well as the other proposed detectors, as shown in 
Figs.~\ref{fig:act} and~\ref{fig:acthm}.

The Whipple detector is located in the northern hemisphere,
and the most likely sites for the VERITAS and HESS arrays are also
in the northern hemisphere (although a southern hemisphere site for the
HESS array is being considered).  Since the galactic center is a
southern source,
it transits at large zenith angles for a northern hemisphere observatory
(e.g., around 61$^\circ$ at the Whipple site).  For such observations at large 
zenith angles, the point of maximum shower development is more distant 
than it would be for showers with normal incidence, and
the Cherenkov light pool spreads out over a larger area.  This has the
advantage of increasing the effective area 
but the disadvantage of raising the energy threshold.  For observations
of the galactic center at the Whipple site, this results in an increase
in the energy threshold and effective area by roughly a factor of
5~\cite{Krennrich}. 
For this
reason we do not consider here the CAT telescope which is located at 
a latitude of $42^\circ$N~\cite{Barrau} 
or the HEGRA array~\cite{Bulian} which will eventually achieve
an energy threshold of 500~GeV at the zenith but a threshold
of 2.5~TeV for the galactic
center, since in both cases their energy thresholds are too high to
be sensitive to the majority of parameter space.
The 3.8m CANGAROO telescope in the southern hemisphere~\cite{Kifune} 
has an energy threshold of approximately 1~TeV and will also not probe
much of the predicted parameter space.  However, the CANGAROO collaboration
is currently constructing a 10~m telescope which may eventually
be part of an array of telescopes.  This would prove to be one of the most
important instruments for galactic center observations.  Since the details
of this proposal are not well established, we consider here a generic 
``southern array'' which has roughly the
characteristics of VERITAS.  Since a source spends the majority of its
time near transit, for such a southern array we do not include the
effect of large zenith angle observations on increasing the effective
area.


\section{Prospectives for the Detection of the Monocromatic $\gamma$ Line with an ACT}\label{sec:actdet}

As we have already pointed out, the detection prospectives for the 
monocromatic $\gamma$ ray line from neutralino annihilation in the 
galactic halo critically depend on the profile which describes the 
dark matter distribution. This is in particular evident for detectors 
like ACTs which have a small angular acceptance. If a large fraction of 
the total flux emitted is concentrated in a tiny region of the sky, the 
galactic center, whose coordinates are known with sufficient accuracy, an 
ACT can be the ideal instrument for detecting neutralino dark matter. This 
is true for singular halo profiles. We will show that with the NFW 
profile, which is not the most singular profile among those that are 
currently being studied, a signal might be seen already in the next 
generation of ACT detectors.

In Fig.~\ref{fig:act} (a) and (b) we show the results of the $2\gamma$ 
and $Z\gamma$ line fluxes for the supersymmetric models considered in 
Section~\ref{sec:mssm} and for the NFW profile which gives the maximal 
value of $\langle\,J\left(0\right)\,\rangle\,(10^{-5}\,sr)$ in 
Fig.~\ref{fig:jd}. We have chosen such a small solid angle, which 
corresponds to the minimal angular acceptance for the observation of a 
point source for an ACT, because for a singular halo profile the ratio 
signal to square root of the background 
increases going to smaller angular acceptances. In 
the figures we draw also the sensitivity curves for each of the ACT 
detectors considered in preavious section. As can be seen, a significant
 number of supersymmetric models could give a signal which exceeds the 
sensitivity of the Southern Array and the VERITAS or HESS detector. The 
detector in the southern emisphere, having a low threshold (around 50 GeV), 
explores a region of the MSSM parameter space which may be accessible to 
other detection methods, in particular to future accelerator experiments. 
VERITAS or HESS have a higher sensitivity for heavier neutralino masses, 
in an interval which will be hardly accessible to direct or other indirect 
 detection methods.

As the energy resolution of the typical ACT is not better than around 10\%, 
in the high mass rage the two monochcromatic $2\gamma$ and $Z\gamma$ lines, respectively at the energy $E_{\gamma} \simeq M_{\chi}$ and 
$E_{\gamma} \simeq M_{\chi} (1 - M_Z^2/4 M_{\chi}^2)$, are not resolvable. In 
Fig.~\ref{fig:acthm} we sum the contributions to the flux from the two annihilation processes. As we have pointed out in Section~\ref{sec:mssm}, a very interesting feature is that due to $Z\gamma$ annihilation line, the values for the predicted flux are concentrated in a quite narrow band and this clearly enhances the possibility of a discovery.


\section{The GLAST satellite detector}\label{sec:glast}

In the existing EGRET, and the planned GLAST
satellite experiment, high energy gamma-rays interact to form
an electron positron pair in a distributed converter/tracker layer.
The measured direction of electron and positron in the
pair (which improves with increasing energy) 
and the subsequent electromagnetic cascade
in the calorimeter layer give the arrival direction and energy of each
gamma-ray.  
An anticoincidence shield effectively rejects background cosmic-ray nuclei
and electrons, and the sensitivity of these detectors is limited only
by counting statistics and by the diffuse gamma-ray background.

For an energy resolution of 4\%, an angular resolution of 0.1$^\circ$
(at $E_\gamma > 10$~GeV)
it is clear that the limit given in Eqs.\,~(\ref{eq:sig}) and
(\ref{eq:min})
 applies for the
point source sensitivity, and that such observations 
are counting-statistics limited.  Given this fact it is
advantageous to change the observing strategy and to 
relax the angular acceptance window in order that the number of collected
gamma-rays is increased above the threshold for detection (around 10 events).

The estimate of the diffuse gamma ray background is then crucial to give 
a prediction for a signal in a satellite based experiment (for ground-based
experiments the  background from misidentified protons is generally higher). 
The 
diffuse emission of gamma rays in the galaxy is thought to be due 
mainly to cosmic-ray protons and electrons interacting with 
the interstellar medium, and has an overall enhancement towards 
the inner galaxy and the galactic disk. Given that the galaxy is 
essentially transparent to gamma rays up to energies of about 100 
TeV, it is possible to derive the gamma ray spectrum in the energy 
region of our interest from interstellar mass models of 
the galaxy ~\cite{bersch,strong,salati}. This is however beyond the 
scope of the paper, we will limit ourselves to an order of magnitude 
estimate which can be obtained extrapolating from existing data at 
lower energies.

The EGRET experiment has mapped 
the diffuse gamma ray emission up to about 10 GeV. We assume a power 
law fall-off in energy in the form:
\beq
\frac{dN(E_{\gamma},l,b)}{dE_{\gamma}} = N_0(l,b) \left( \frac{E_{\gamma}}{1\, \rm{GeV}}\right)^{\alpha}\;10^{-6}\; \rm{cm}^{-2}\,\rm{s}^{-1}\,\rm{GeV}^{-1}\,\rm{sr}^{-1}\label{eq:bkg}
\eeq
where we have supposed that the normalization factor $N_0$  depends 
only on the interstellar matter distribution and is a 
function  of the galactic coordinates $(l,b)$ only.  
$N_0$ has been fixed using the EGRET data at 1 GeV published 
in Ref.~\cite{hunter}, in the simple functional form
\beqa 
N_0(l,b)&=\frac{\mbox{\normalsize $85.5$}}{\sqrt{1+\left(l/35\right)^2}\;\sqrt{1+\left(b/(1.1+|l|\,0.022)\right)^2}} \,+\,0.5 & \rm{if}\; |l|\,
\geq 30^{\circ} \nonumber \\
&=\frac{\mbox{\normalsize $85.5$}}{\sqrt{1+\left(l/35\right)^2}\;\sqrt{1+\left(b/1.8\right)^2}}\,+\,0.5\;\;\;\;\;\;\;\;\;\;\;\;\;\;\; & \rm{if} \; |l|\,\leq 30^{\circ}\label{eq:bkg1}
\eeqa
where the longitude $l$ is assumed to vary in the intervall $-180^{\circ}
\leq l \leq 180^{\circ}$ and the latitude $b$ in $-90^{\circ}\leq b 
\leq 90^{\circ}$. The parametrization is in resonable agreement with the 
data at least in the region towards the galactic centre, in which we will 
use it. The slope parameter $\alpha$ may in principle depend on both energy
and the galactic coordinates. We will, however, adopt $\alpha=-2.7$ in lack
of data for the energies of interest to us.

\subsection{Acceptance and energy resolution}

The proposed GLAST satellite detector 
will have a much smaller area (around
1 $m^2$) than  an ACT. However, the solid angle coverage will be
of the order of one steradian and the energy resolution will be excellent,
at the per cent level for gamma rays that enter on the side of the detector
and thus pass many radiation lengths of the electromagnetic calorimeter.

As an example \cite{francke}, the sensitivity of one version of 
the GLAST calorimeter, consisting of 25
towers (each 32 by 32 cm$^2$) of 10 radition lengths of CsI, has been
 simulated using the GEANT program. The geometric
 acceptance of the instrument was calculated for all gamma rays which
 fulfil that a cylinder around the electromagnetic shower of more than
 18 radiation lengths and a radius of at least two Moliere radii is
 fully contained within the instrument, giving an energy resolution of
 better than 1.5\% for gamma rays with energies higher than 50 GeV. 
It was assumed that the diffuse gamma
 rays enter the instrument from any direction in the upper hemisphere.
 (as the lower hemisphere will be completely screened by the Earth). 
This results in a geometric acceptance of 2.1 m$^2$sr. 

 Taking into account the screening by the Earth of the galactic 
center region half of the 
orbiting time, the useful geometric acceptance in a 1 sr cone
towards the galactic center is 0.18 m$^2$sr. 
This is the geometric acceptance both for a
 monoenergetic neutralino annihilation signal from a region around the
 center of the galaxy as well as the diffuse gamma ray background,
 since background photons arriving from other directions can be
 discriminated by the direction of incidence. 

A similar analysis has to
our knowledge not been performed for the alternative design SIFTER, 
but it appears that
there the somewhat larger effective area for a given weight of the detector
can give a similar improvement of the $\gamma$ ray line discovery potential
\cite{pendelton}.

\begin{figure}
 \centering
 \mbox{\subfigure{\epsfig{file=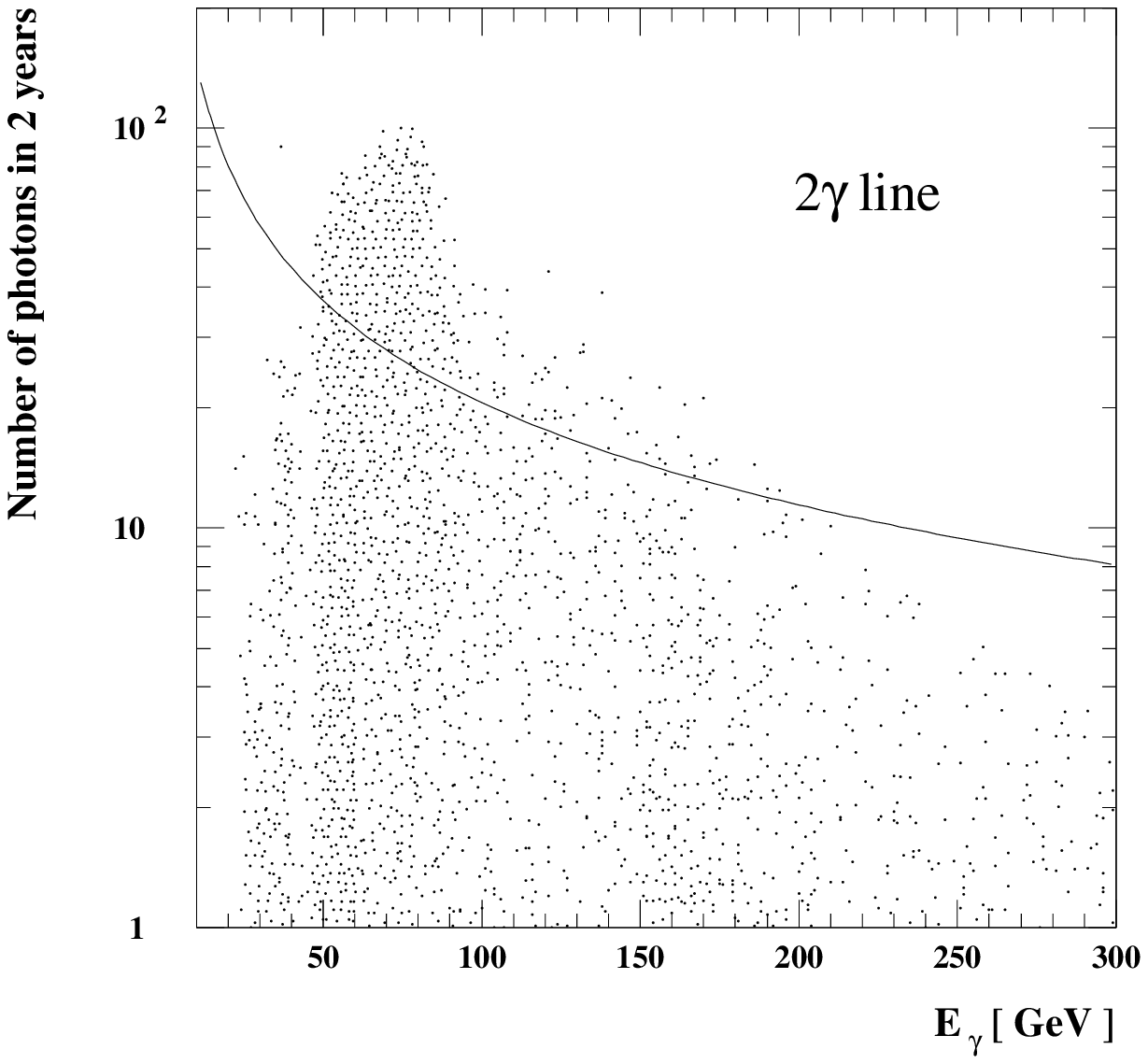,width=7cm}} \quad
       \subfigure{\epsfig{file=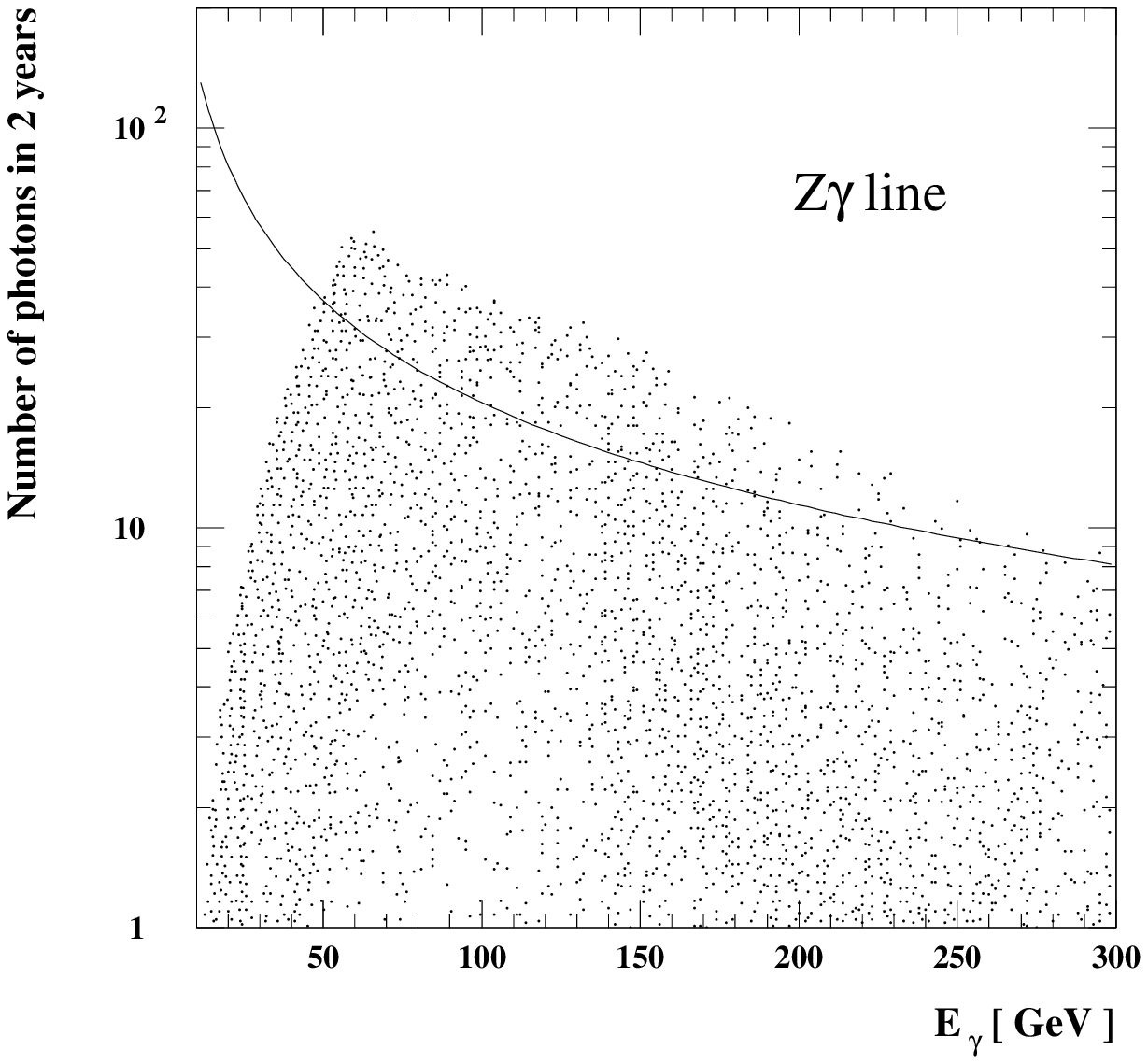,width=7cm}}}
\caption[]{The number of events expected in GLAST from a 1 sr cone encompassing
 the galactic center, assuming a 2 year exposure and calorimetry as described
in the text, for the $2\gamma$ (on the left) and the $Z\gamma$ 
annihilation line (on the right). 
The NFW halo profile giving the maximal flux 
has been assumed. The solid line shows the number of
events needed to obtain a $5\sigma$ detection over the background as estimated
from EGRET data.} 
\label{fig:glast}
\end{figure}

We display in Fig.\,~\ref{fig:glast} (a) and (b) the results of the 
$2\gamma$ and $Z\gamma$  line flux from Section \ref{sec:mssm}
using the GLAST parameters just discussed, 
and assuming a two-year exposure. Also shown
is the curve giving the minimum number of events needed to observe an
effect at the $5\sigma$ level, using 
Eq.\,(~\ref{eq:bkg}) for the galactic diffuse $\gamma$-ray background flux.
The  NFW halo model corresponding to the maximal rate was assumed for the halo.

As can be seen, a fair fraction of our set of MSSM points can be probed under 
these circumstances. An advantage with a detector of large acceptance is that
``clumps'' of dark matter of relatively small angular  size could still
be detected. Since the annihilation flux is proportional to the square
of the dark matter density, such enhancements would appear as bright spots
on the $\gamma$ ray sky, without optical counterpart.
Another advantage with a satellite experiment is the excellent energy 
resolution possible, which may enable a separation of the two lines from
$2\gamma$ and $Z\gamma$ respectively. If both lines were to be
observed, a comparison of line strengths would
give interesting information on the supersymmetric model.

If the relative energy resolution is $\epsilon = \sigma_E/E$, 
the two lines are separable provided that 
\beq
M_\chi\lsim {M_Z\over \sqrt{4\epsilon}},
\eeq
which for $\epsilon=0.015$ gives $M_\chi\lsim 370$ GeV.

\section{Continuum Gamma Rays}\label{sec:cont}

Besides the gamma ray line coming from the $\gamma\gamma$ and $Z\gamma$
final states, there may also be produced a continuum $\gamma$ spectrum 
mainly from decaying $\pi^0$ mesons created in the fragmentation of quarks
\cite{bengtsson,berezinsky,chardonnet}. 
 The general drawback of the continuum 
$\gamma$ spectrum is that in contrast to the $\gamma$ lines 
it lacks distinctive features which can exclude other more mundane sources
of production if a signal is believed to be found. On the other hand,
the number of photons from this continuum source can be much higher 
than from the line processes.

In fact, should a $\gamma$ ray line be found, by necessity there has
to be a continuum signal also, although in some cases severely affected
by the galactic diffuse background. We therefore investigate
what can be expected for the higgsino-like high-mass sample, since previous
treatments were mainly done for low-mass neutralinos. 

\begin{figure}[htb]
 \epsfig{figure=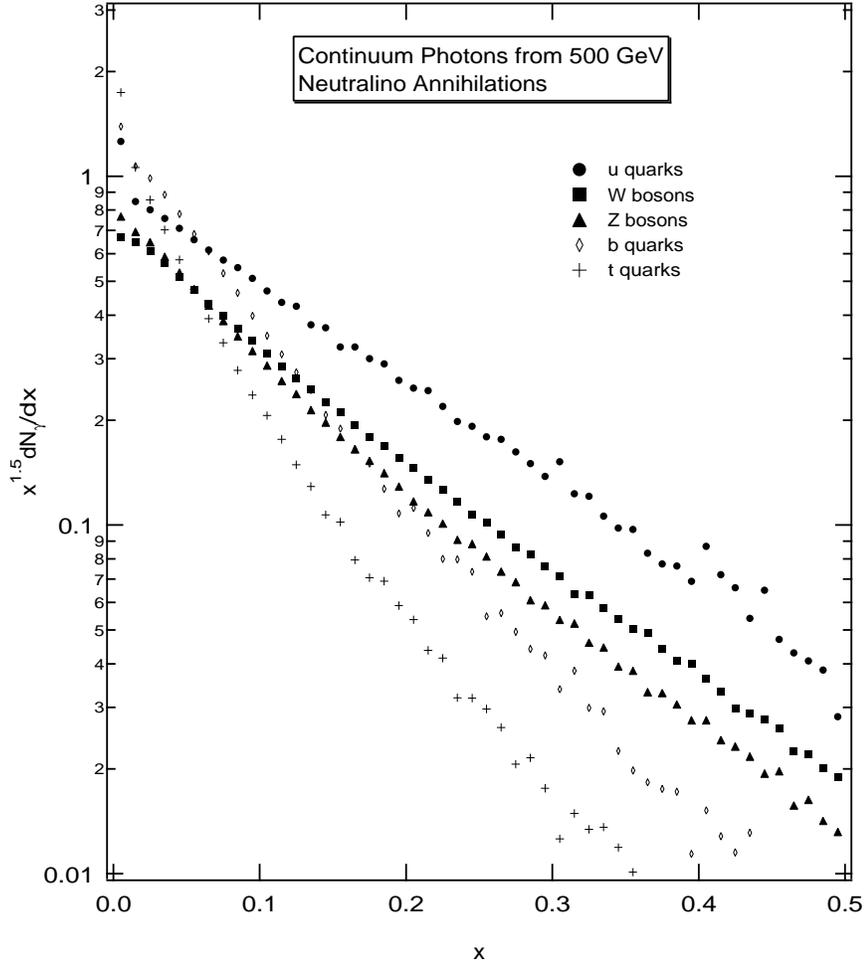,width=12cm}
\caption[]{Continuum photons from 500 GeV neutralino annihilations. 
The rate is given as $x^{1.5}dN_{\gamma}/dx$ per final state pair for
$u$ quarks (filled circles), $W$ bosons (filled squares), $Z$ bosons 
(filled triangles), $b$ quarks (open diamonds) and $t$ quarks 
(crosses). The scaling variable $x$ is defined as 
$x=E_{\gamma}/m_{\chi}$, where $E_{\gamma}$ is the photon energy and 
$2m_{\chi}$ the total energy of the final state in the annihilation.
Note that for a realistic neutralino of this high mass, the 
annihilation into $u$ quarks, normalized to unity in the figure, will 
generally be much suppressed.}
\label{fig:pythia}
\end{figure}

For our high-mass sample, the 
main annihilation mode is into a $WW$ or $ZZ$ pair. We have simulated 
the continuum photon spectrum from these and $q\bar q$ final states 
through the 
PYTHIA 6.113 Monte Carlo code \cite{sjostrand}. Introducing the 
scaling variable $x=E_{\gamma}/M_{\chi}$, we find empirically that 
the quantity 
$x^{1.5}dN_{\gamma}/dx$ is well described by an exponential in the 
variable $x$, as shown in Fig.\,~\ref{fig:pythia} for $M_{\chi}=500$ 
GeV.

\begin{figure}[htb]
 \epsfig{figure=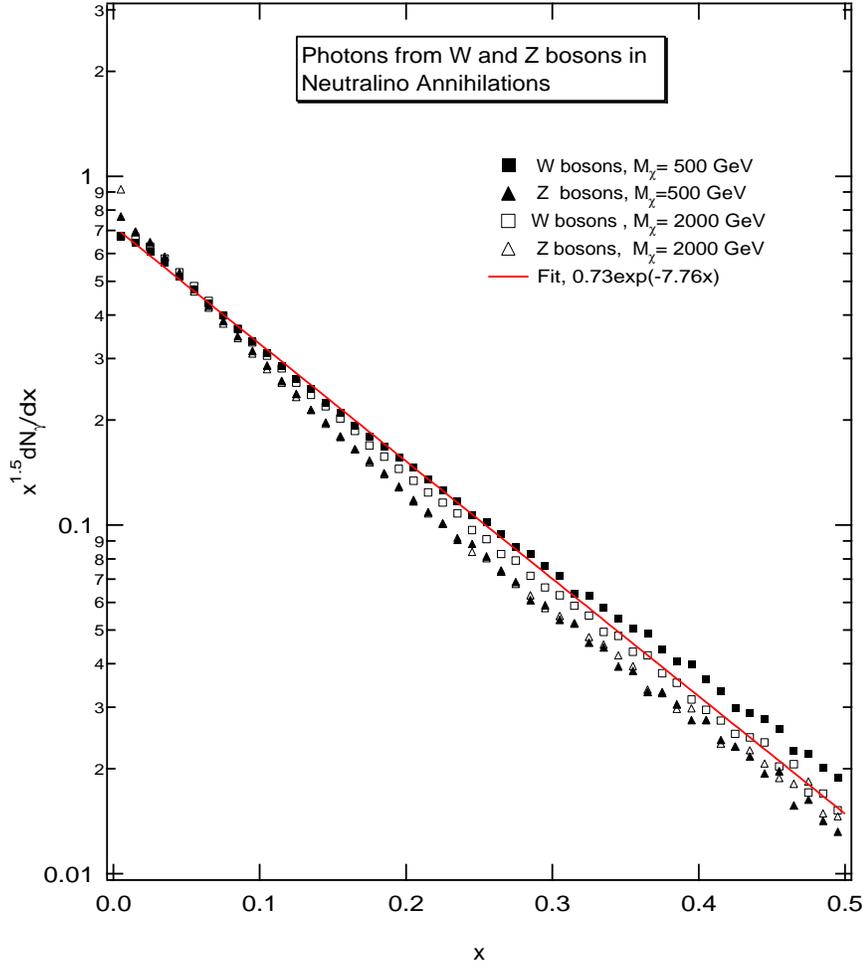,width=12cm}
\caption[]{Continuum photons from $W$ and $Z$ bosons in
 in high-mass neutralino (in practice, mostly higgsino) annihilations.
Pythia \protect\cite{sjostrand} results for 500 GeV neutralinos are shown
with filled symbols, for 2000 GeV with open symbols. The line gives the
analytic fit according to Eq.\,~(\ref{eq:scaling}).}
\label{fig:pythia2}
\end{figure}

As can be seen, $t$ and $b$ quarks give a quite soft spectrum. Since 
$u$ quarks (as well as $s$ and $d$ quarks) have a very small branching 
ratio for these high mass neutralinos, the dominant source of photons 
above $x\simeq 0.1$ will be $WW$ and $ZZ$. Both of these final states 
show a very simple scaling behavior when increasing the energy (i.e. 
the neutralino mass). This is clearly shown in 
Fig.\,~\ref{fig:pythia2}, where photons from both $WW$ and $ZZ$ at 
$M_{\chi}$ of both 500 and 2000 GeV can be described to the level of 
accuracy required for our purpose by the distribution
\beq
{dN_{\gamma}\over dx}={M_\chi dN_\gamma\over dE_\gamma}=
{0.73\over x^{1.5}}e^{-7.8x}.\label{eq:scaling}
\eeq

The models in our sample with $M_{\chi}>300$ GeV with the highest rate 
of $WW$ and $ZZ$ production have a rate given approximately by
\beq
(\sigma v)_{WW+ZZ}^{max}\simeq 5\cdot 10^{-26} \left({300\ {\rm 
GeV}\over M_{\chi}}\right)^2\ \ {\rm cm}^3{\rm s}^{-1}.
\eeq
The observed differential flux at the Earth is then given by
\beqa
\phi_{\gamma}^{{\rm continuum}}(E_{\gamma})&\simeq& 1.2\cdot 10^{-10}\left({300\ {\rm 
GeV}\over M_{\chi}}\right)^4{dN_{\gamma}\over dE_{\gamma}}\cdot \nonumber \\
&&\;\;\;\;\;\;\cdot J(\Psi)\; {\rm cm}^{-2}{\rm s}^{-1}{\rm sr}^{-1}{\rm GeV}^{-1}\; ,
\label{eq:cont1}
\eeqa
where $J(\Psi)$ is defined in Eq.\,~(\ref{eq:jpsi}). 

\begin{figure}[htb]
 \epsfig{figure=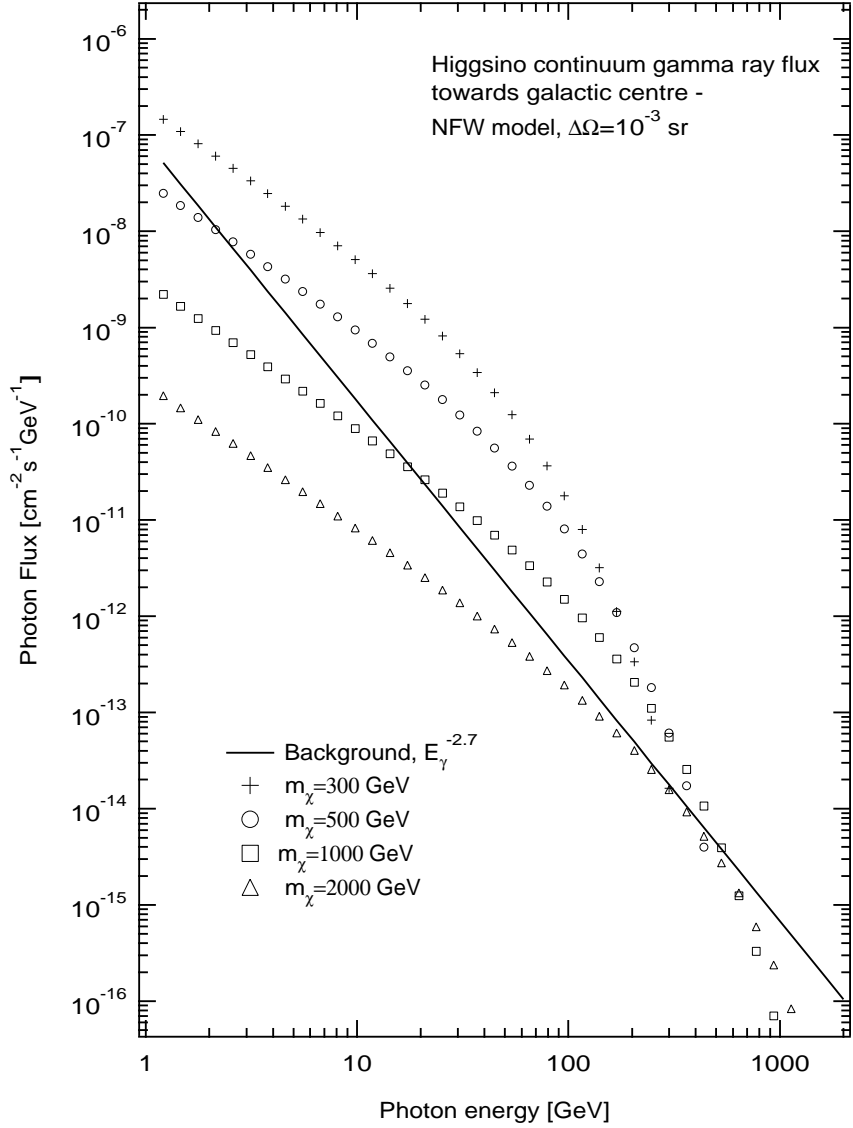,width=12cm}
\caption[]{Continuum photons from the direction of the galactic centre 
originating from $W$ and $Z$ bosons in
 neutralino annihilations, for $m_{\chi}=300, 500, 1000$ and $2000$ 
 GeV. The NFW halo profile giving the maximal signal has been assumed, 
 and an angular integration over $10^{-3}$ sr performed. The 
 background flux (solid line) is that predicted by Eq.\,~(\ref{eq:bkg}).}
\label{fig:cont1}
\end{figure}

Using the NFW halo profile which gives the highest rate, and integrating 
over $\Delta\Omega=10^{-3}$ sr towards the galactic center,
this gives the results shown in Fig.\,~\ref{fig:cont1} for 
$m_{\chi}=300, 500, 1000$ and $2000$ GeV. The galactic background curve 
is the estimate Eq.\,(~\ref{eq:bkg}), which for $b,l=0$ gives 
\beq
\phi_{bkg}^{g.c.}(E_{\gamma})\sim 6\cdot 10^{-5}\left(\frac{E_{\gamma}}
{1\ {\rm GeV}}\right)^{-2.7} \rm{cm}^{-2}\,\rm{s}^{-1}\,\rm{sr}^{-1}
\,\rm{GeV}^{-1}\; .
\eeq

\begin{figure}[htb]
 \epsfig{figure=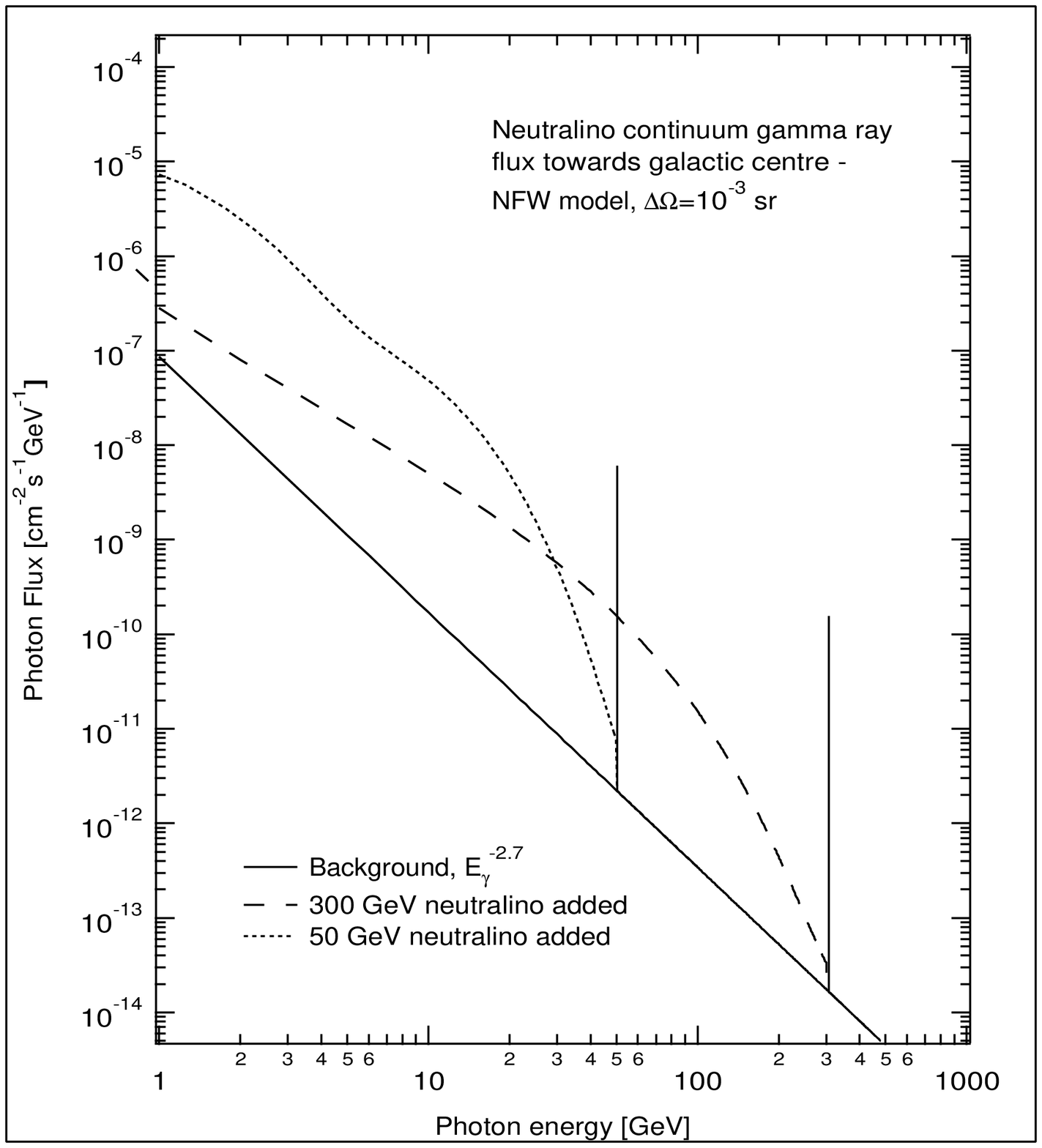,width=12cm}
\caption[]{Total photon spectrum from the direction of the galactic centre 
originating from $W$ and $Z$ bosons in
 neutralino annihilations, for $300$ GeV, and from $b$ quarks for $50$ 
 GeV. The NFW halo profile giving the maximal signal has been assumed, 
 and an angular integration over $10^{-3}$ sr performed. The 
 background flux is that predicted by Eq.\,~(\ref{eq:bkg}).} In addition,
the maximal $\gamma$ line strength found in our sample is displayed for
these two masses, assuming a relative line width of $10^{-3}$.
\label{fig:cont2}
\end{figure}

We see that below a mass of 1000 GeV, this additional component of 
the photon spectrum is larger than the estimated background over some 
energy range, and  could thus
be observable, see Fig.\,~\ref{fig:cont2}, where also the results for
a 50 GeV neutralino (annihilating mainly to $b\bar b$) pairs are shown, using
the parametrisation of \cite{bengtsson} of the  $\gamma$ spectrum from
$b$ quarks. 
As can be seen, at least the 50 GeV candidate would appear to be excluded 
already by EGRET data, were it not for the uncertainties related to 
the halo parameters. 
The combined features of
a break in the spectrum and an angular distribution that does not follow that
of the disk and bulge of the galaxy could make this type of 
signal distinguishable
from the background, although it may be difficult to exclude other sources
of $\gamma$ rays giving similar features. In fact, present EGRET observations 
are not inconsistent with a continuum spectrum originating from dark matter
annihilations, but other explanations are possible as well \cite{dixon}.

It is interesting to note in Fig.\,~\ref{fig:cont2} that a 300 GeV higgsino
would cause a flatter spectrum than the ``canonical'' $E_\gamma^{-2.7}$
in the 1 - 100 GeV range. It is intriguing that a flatter spectrum is
indeed observed towards the galactic centre, something which has turned
out difficult to reproduce in conventional models \cite{mori}. Before detection
of a line signal it will, however, not be possible to rule out other
sources of high energy $\gamma$ rays. In any case, the hitherto unexplored
energy band between 100 and 300 GeV will be of crucial importance to verify
or rule out a supersymmetric component of the $\gamma$ ray spectrum.

The higher-mass neutralinos (above 1 TeV) do not give as useful
a continuum signal because of the rapid decrease with $m_\chi$ of the 
signal (see Eq.\,~(\ref{eq:cont1})), whereas the $\gamma$ line remains
a viable signature up to that region.

\section{Discussion}\label{sec:disc}

To conclude, we have shown that the new high energy cosmic 
$\gamma$-ray detectors, both ground- and space-based will have 
an interesting opportunity to search for signals of dark matter 
paticle annihilations in the Milky Way halo. Indeed, existing data
from EGRET may already rule out some of the low-mass MSSM models
if the halo is of the most singular form studied here.

Even neutralinos as heavy as several hundred GeV could give continuum
$\gamma$s with an observable  flatter spectrum than the $E_\gamma^{-2.7}$
expected in the simplest models of diffuse galactic $\gamma$ rays, in
the direction towards the galactic center. The absence of a clear
signature may, however, make an eventual 
claim of detection of a supersymmetric  signal difficult to defend. 

The
only reliable $\gamma$-ray signals indicating directly the existence
of slowly moving, heavy neutral particles in the Milky Way halo seem
to be the $\gamma$-ray lines produced by $\chi\chi\to\gamma\gamma$
and $\chi\chi\to Z\gamma$ annihilations. These lines, of intrinsic
relative width of the order of $10^{-3}$, have no plausible astrophysical
background whatsoever and would constitute a ``smoking gun'' of
supersymmetric dark matter.

As we have shown, the larger rates for the line processes found in
the recent first full one-loop calculations in the MSSM, together with the
central enhancement indicated in simulations of cold dark matter halos,
make these processes accessible to the next generation of $\gamma$-ray
detectors over a fair range of parameters. In particular, if the
neutralino is the lightest supersymmetric particle and is 
heavier than around 1 TeV, detection of 
the $\gamma$ line may well be the only way to discover supersymmetry
in the foreseeable future, as all other methods (accelerator, direct
detection and indirect neutrino detection) are discouragingly far from
the required sensitivity.


\vskip 2.0 cm
{\bf\Large Acknowledgments}
\vskip .2cm
We are grateful to J. Eds\"o and P. Gondolo for collaboration on
the numerical programs used.
This work was supported with computing resources by the Swedish Council 
for High Performance Computing (HPDR) and Parallelldatorcentrum (PDC),
Royal Institute of Technology, Stockholm. We thank J. Edsj\"o 
for very valuable help with  supercomputer runs.
L.B. wishes to thank the 
Swedish Natural Science Research Council (NFR) for support.
We are grateful to P. Carlson and T. Francke
for helpful discussions on the GLAST calorimeter, and G. Pendelton for 
information on SIFTER.

\pagebreak

\begin{thebibliography}{99}

\bibitem{olive} See, e.g., K.A. Olive, astro-ph/9712160, to appear in
Proc. TAUP97, Nucl. Phys. Suppl., A. Bottino, A. di Credico and 
P. Monacelli (eds.), in press.  
\bibitem{lbetaup} For a recent review, see, e.g., L. Bergstr\"om, 
astro-ph/9712179, to appear in
Proc. TAUP97, Nucl. Phys. Suppl., A. Bottino, A. di Credico and 
P. Monacelli (eds.), in press.      
\bibitem{jkg} G. Jungman, M. Kamionkowski and K. Griest, Phys. Rep.
{\bf 267} (1996) 195.
\bibitem{lp} L.~Bergstr{\"o}m and P.~Ullio, Nucl.\ Phys.\ {\bf B504} 
(1997) 27; see also Z. Bern, P. Gondolo and M. Perelstein, 
Phys. Lett. {\bf B411} (1997) 86.
\bibitem{lp2} P.~Ullio and L.~Bergstr{\"o}m, Phys. Rev. D, in press.
\bibitem{neutrinos} J. Silk, K.A. Olive and M. Srednicki, Phys. Rev. Lett.
{\bf 55} (1985) 257; 
T.K. Gaisser, G. Steigman and  S. Tilav, Phys.Rev. {\bf D34} (1986) 2206;
S. Ritz and D. Seckel,   Nucl.Phys. {\bf B304} (1988) 877;
G.F. Giudice and E. Roulet, Nucl.Phys. {\bf B316} (1989) 429;
G. Jungman and M. Kamionkowski, Phys.Rev. {\bf D51} (1995) 328;
L. Bergstr\"om, J. Edsj\"o and P. Gondolo, Phys.Rev. {\bf D55} (1997) 1765.
\bibitem{oldlines}L. Bergstr\"om and H. Snellman, Phys. Rev. {\bf D37} 
(1988) 3737; S. Rudaz, Phys.Rev. {\bf D39} (1989) 3549; G.F. Giudice and 
K. Griest, Phys.Rev. {\bf D40} (1989) 2549; A. Bouquet, P. Salati and
J. Silk, Phys.Rev. {\bf D40} (1989) 3168; 
F. Stecker and A. Tylka, Astrophys. J. {\bf 336} (1989) L51;
V. Berezinsky, A. Bottino and
V. de Alfaro, Phys.Lett. {\bf B274} (1992) 122;
M. Urban et al.,Phys.Lett. {\bf B293} (1992) 149;
L. Bergstr\"om and J. Kaplan, Astropart.Phys. {\bf 2} (1994) 261;
G. Jungman and M. Kamionkowski, Phys.Rev. {\bf D51} (1995) 3121.
\bibitem{carlberg} R. Carlberg, Astrophys. J. {\bf 433} (1994) 468.
\bibitem{navarro}J.F. Navarro, C.S. Frenk and S.D.M. White, Astrophys. J.
{\bf 462} (1996) 563.
\bibitem{kravtsov} A. V. Kravtsov et al., astro-ph/9708176 (1997).
\bibitem{moore} B. Moore et al., astro-ph/9709051, Astrophys. J. Lett.,
submitted.
\bibitem{bs} A. Burkert and J. Silk, astro-ph/9707343 (1997).
\bibitem{bg} L. Bergstr\"om and P. Gondolo, Astropart. Phys. {\bf 5}
(1996) 183.
\bibitem{joakim} 
J.~Edsj{\"o}, Aspects of Neutrino Detection of Neutralino Dark Matter 
(Uppsala University thesis, Uppsala, 1997), hep-ph/9704384. 
\bibitem{paolograc} P. Gondolo and G. Gelmini, Nucl. Phys. {\bf B360}
(1991) 145.
\bibitem{joakimpaolo} J. Edsj\"o and P. Gondolo, Phys. Rev. {\bf D56} 
(1997) 1879.
\bibitem{binney} W. Dehnen and J. Binney, astro-ph/9612059 (1997).
\bibitem{bere} V.S.~Berezinsky, A.V.~Gurevich and K.P.~Zybin, 
Phys. Lett. {\bf B294} (1992) 221.
\bibitem{flores} R.A. Flores and J.R. Primack, Astrophys. J. {\bf 427} (1994)
L1.
\bibitem{makino} T. Fukushige and J. Makino, Astrophys. J. {\bf 487} (1997) L9.
\bibitem{evans}N.W. Evans and J.L. Collett, astro-ph/9702085.
\bibitem{kochanek} C.S. Kochanek, Astrophys. J. {\bf 457} (1996) 228.
\bibitem{lin} D.N.C. Lin, B.F.~Jones and A.R.~Klemola, Astrophys. J. 
{\bf 439} (1995) 652.
\bibitem{kerr} F.J. Kerr and D. Lynden-Bell, MNRAS {\bf 221} (1986) 1023.
\bibitem{reid} M.J. Reid, ARA\&A {\bf 31} (1993) 345.
\bibitem{oll} R.P. Olling and M.R. Merrifield, aspro-ph/9711157, to appear in proceedings of the Workshop on Galactic Halos, Santa Cruz, August 1997 (ASP conference Series).
\bibitem{kui} K. Kuijken and G. Gilmore Astrophys. J. {\bf 367} (1991) L9.
\bibitem{gould} A. Gould, MNRAS {\bf 244} (1990) 25.
\bibitem{schneider} P. Schneider, J. Ehlers and E.E. Falco, 
{\em Gravitational Lenses}, Springer-Verlag, Berlin, 1992.
\bibitem{Cawley} M.F.~Cawley et al., Exper.\ Astr. {\bf 1} (1990) 173. 
\bibitem{Weekes} T.C.~Weekes et al., in Proc. of the 25th ICRC, 5,(1997) 173.
\bibitem{goret} P. Goret et al., astro-ph/9710260.
\bibitem{Aharonian} F.A.~Aharonian, W.~Hofmann, A.K.~Konopelko and H.J.~V\"olk, Astroparticle Physics {\bf 6} (1997) 343. 
\bibitem{Barrau} A.Barrau et al. 1997, astro-ph/9705249.
\bibitem{Reynolds} P.T.~Reynolds, C.W.~Akerlof and M.F.~Cawley, 
ApJ {\bf 404} (1993) 206.  
\bibitem{Buckley} J.H.~Buckley et al., A\&{A} {\bf 329} (1998) 63.
\bibitem{hunter} S.D. Hunter et al., Ap. J. {\bf 481} (1997) 205.
\bibitem{longair} M.S.~Longair, High Energy Astrophysics, Cambridge
U.~Press, Cambridge (1992).
\bibitem{Krennrich} F.~Krennrich, ApJ {\bf 481} (1997) 758.  
\bibitem{Bulian} N.Bulian et al. 1997, astro-ph/9712063.
\bibitem{Kifune} T. Kifune et al., in Proc. of the 4th Compton
Symposium, Williamsburg (1997), in press, astro-ph/9707001.
\bibitem{bersch} D.L. Bertsch et al., Ap. J. {\bf 416} (1993) 587.
\bibitem{strong} I.V. Mosalenko and A.W. Strong, Ap. J. {\bf 493} 
(1998) in press.
\bibitem{salati} P. Chardonnet et al, Ap. J. {\bf 454} (1995) 774.
\bibitem{francke} L. Bergstr\"om, P. Carlson and T. Francke, unpublished.
\bibitem{pendelton} G. Pendelton, private communication.
\bibitem{bengtsson} H.-U. Bengtsson, P. Salati and J. Silk, Nucl Phys.
{\bf B346} (1990) 129.
\bibitem{berezinsky} V. Berezinsky, A. Bottino and  G. Mignola,
  Phys. Lett. {\bf B325} (1994) 136.
\bibitem{chardonnet} P. Chardonnet et al., Astrophys. J. {\bf 454} (1995) 774.
\bibitem{sjostrand} T. Sj\"ostrand, Comp. Phys. Comm. {\bf 82} (1994) 74.
\bibitem{dixon} D. Dixon et al., as quoted in press release, 
http://tigre.ucr.edu/halo/halo.html.
\bibitem{mori} M.Mori, Astrophys. J. {\bf 478} (1997) 214.
\end{thebibliography}
\end{document}